\documentclass[12pt,a4paper]{article}
\input epsf
\usepackage{amsmath}
\usepackage{amssymb}
\usepackage{graphicx}

\newcommand{\atopfrac}[2]{\genfrac{}{}{0pt}{}{#1}{#2}}
\newcommand{\sfrac}[2]{{\textstyle\frac{#1}{#2}}}
\newcommand{\half}{\sfrac{1}{2}}
\newcommand{\quarter}{\sfrac{1}{4}}

\newcommand{\alg}[1]{\mathfrak{#1}}

\newcommand{\indups}[1]{_{\mathrm{\scriptscriptstyle #1}}}
\newcommand{\gym}{g\indups{YM}}

\newcommand{\Tr}{{\rm Tr \,}}
\newcommand{\up}{\uparrow}
\newcommand{\dn}{\downarrow}
\newcommand{\updn}{\updownarrow}
\newcommand{\hole}{\circ}
\newcommand{\cO}{{\cal O}}
\newcommand{\cN}{{\cal N}}

\begin{document}
\thispagestyle{empty}
\begin{flushright}
{\sc\footnotesize hep-th/0512077} \\
{\sc\footnotesize AEI-2005-164}\\
{\sc\footnotesize SPhT-T05/190}\\
{\sc \footnotesize NSF-KITP-05-84}
\end{flushright}
\vspace{1cm}
\setcounter{footnote}{0}
\begin{center}
{\Large{\bf Planar \mathversion{bold}${\cal N}=4$ Gauge Theory 
and the Hubbard Model \par}}\vspace{20mm}
{\sc Adam Rej$^a$, Didina Serban$^{b,c}$ and Matthias Staudacher$^{a,c}$} 
\\[7mm]
$^a${\it Max-Planck-Institut f\"ur Gravitationsphysik\\
     Albert-Einstein-Institut \\
     Am M\"uhlenberg 1, D-14476 Potsdam, Germany}\\ [2mm]
$^b${\it Service de Physique Th\'eorique, CNRS-URA 2306
\\
C.E.A.-Saclay \\
F-91191 Gif-sur-Yvette, France}\\[1mm]
{\it and} \\ [1mm]
$^c${\it Kavli Institute for Theoretical Physics,
University of California\\
Santa Barbara, CA 93106 USA} \\ [2mm]
{\tt Adam.Rej@aei.mpg.de, serban@spht.saclay.cea.fr, matthias@aei.mpg.de} 
\\[10mm]
{\sc Abstract}\\[2mm]
\end{center}

\noindent{ 
Recently it was established that a certain integrable {\it long-range} 
spin chain describes the dilatation operator of ${\cal N}=4$ gauge 
theory in the $\alg{su}(2)$ sector to at least three-loop order, 
while exhibiting BMN scaling to all orders in perturbation theory. 
Here we identify this spin chain as an approximation to an
integrable {\it short-ranged} 
model of strongly correlated electrons: The Hubbard model. 
}

\newpage

\setcounter{page}{1}

\section{Introduction}
\label{sec:intro}

Recently it was discovered that the planar one-loop dilatation operator of
supersymmetric ${\cal N}=4$ gauge theory is completely integrable
\cite{MZ,BS1}. This means that its spectrum may be exactly determined
in the form of a set of non-linear Bethe equations. Evidence was
found that this integrability is preserved beyond the one-loop
approximation, and it was conjectured that the dilatation operator
might be integrable to all orders in perturbation theory \cite{BKS}.
Given the usually benign, analytic nature of planar perturbation theory,
one may then even hope for the theory's complete large $N$ integrability 
at all values of the Yang-Mills coupling constant.

Deriving the dilatation operator from the field theory, and
subsequently demonstrating its integrability, is not easy.
The three-loop planar dilatation operator in the maximally compact
$\alg{su}(2|3)$ sector was found by Beisert, up to two unknown constants, 
by algebraic means in \cite{dynamic}. These constants could later be 
unequivocally fixed from the
results of a solid field theory calculation of Eden, Jarczak and 
Sokatchev \cite{EJS}. This basically completely determines the
planar dilatation operator in this large sector up to three loops.
Its restriction to $\alg{su}(2)$ agrees with the original
conjecture of \cite{BKS}. Three-loop integrability in $\alg{su}(2)$
was then demonstrated in \cite{SS} by embedding the 
dilatation operator into an integrable long-range spin chain
due to Inozemtsev, and a three-loop Bethe ansatz was derived.

The Inozemtsev spin chain exhibited a four-loop breakdown of
BMN scaling \cite{BMN}. This scaling behavior seemed, and still seems, to be
a desirable, albeit unproven, property of perturbative gauge theory. 
Mainly for that reason an alternative long-range spin chain,
differing from the Inozemtsev model at and beyond four loops,
was conjectured to exist in \cite{BDS}. Its construction principles
were an extension of the ones already laid out in \cite{BKS}:
(1) Structural consistency with general features of Yang-Mills
perturbation theory, (2) perturbative integrability and 
(3) qualitative BMN scaling. The model's Hamiltonian
is only known up to five loops, and increases exponentially in
complexity with the loop order. In striking contrast, a very compact
Bethe ansatz may be conjectured for the model and shown to diagonalize
the Hamiltonian to the known, fifth, order. The conjecture reads
\begin{equation}
\label{bds}
e^{i p_kL}=
\prod_{\textstyle\atopfrac{j=1}{j\neq k}}^M
\frac{u_k-u_j +i}{u_k-u_j-i}\, ,
\quad k=1,\ldots, M\, ,
\end{equation}
where the rapidities $u_k=u(p_k)$ are related to the momenta $p_k$ 
through the expression
\begin{equation}
\label{u}
u(p_k) =\frac{1}{2}\cot \frac{p_k}{2}\, \sqrt{1+8g^2\sin^2 \frac{p_k}{2}}\;,
\end{equation}
and the energy should be given by
\begin{equation}
\label{bdseng}
E(g)=-\frac{M}{g^2}+\frac{1}{g^2}\sum_{k=1}^M\,
 \sqrt{1+8g^2\sin^2 \frac{p_k}{2}}\;.
\end{equation}
This Bethe ansatz should yield the anomalous dimensions $\Delta$ of
 $\alg{su}(2)$ operators of the form 
\begin{equation}
\label{ops}
\Tr X^M Z^{L-M}+\ldots\, ,
\quad {\rm where} \quad
\Delta(g)=L+g^2\,E(g)\, 
\quad {\rm with} \quad 
g^2=\frac{\gym^2 N}{8 \pi^2}=\frac{\lambda}{8 \pi^2}\, .
\end{equation}
The dots indicate all possible orderings of the partons $Z$ and $X$ 
inside the trace. This mixing problem is diagonalized by the spin chain
Hamiltonian, where we interpret $Z$ as an up-spin $\uparrow$ and $X$ as a
down spin $\downarrow$. $L$ is the length of the spin chain, and
$M$ the number of magnons $\downarrow$. These are the elementary 
excitations on the ferromagnetic vacuum state 
$|\uparrow \uparrow \ldots \uparrow \uparrow\rangle$ which should
be identified with the gauge theory's BPS state $\Tr Z^L$.
To leading one-loop order the spin chain Hamiltonian coincides with
the famous isotropic nearest-neighbor Heisenberg XXX spin chain
\cite{MZ}, and the corresponding Bethe ansatz is obtained by
taking the $g \rightarrow 0$ limit of \eqref{bds},\eqref{u},\eqref{bdseng}.
See also \cite{thesis} for a detailed explanation of the long-range
spin chain approach to gauge theory.

The higher-loop Bethe ansatz \eqref{bds},\eqref{u},\eqref{bdseng} has many 
intriguing properties \cite{BDS}, and it is suspicious that it
should not have already appeared before in condensed matter theory. 
It is equally curious that the Hamiltonian should be so complicated, see
\cite{BDS}, to the point that it is unknown how to write it down in 
closed form. Finally, and most importantly, the Bethe ansatz is expected 
to break down at wrapping order, i.e.~it is {\it not} believed
to yield the correct anomalous dimensions $\Delta$ starting from
${\cal O}(g^{2 L})$. This suggests that the asymptotic Bethe ansatz 
\eqref{bds},\eqref{u},\eqref{bdseng} is actually
not fully self-consistent at finite $L$ and $g\neq 0$.

Nearly all work in solid state theory on the Heisenberg magnet 
has focused on the antiferromagnetic vacuum and its ``physical'' 
elementary excitations, the spinons. The only notable exceptions
seem to be two articles of Sutherland and of Dhar and Shastry
\cite{ferromagnet}, where it was noticed that
the dynamics of magnons in the ferromagnetic vacuum is far from trivial. 
This was later independently rediscovered and extended in the
${\cal N}=4$ context in \cite{MZ,BMSZ}. In gauge theory the
BPS vacuum is very natural, but it should be stressed that
{\it all} states are important. In particular, it is interesting
to ask what is the state of {\it highest} possible anomalous dimension.
This is precisely the antiferromagnetic vacuum state, where $M=L/2$
and $E(g)$ in \eqref{bdseng} should be maximized. Contrary to the BPS
state $|\uparrow \uparrow \ldots \uparrow \uparrow\rangle$
this state is highly nontrivial, as the N\'eel
state $|\uparrow \downarrow \uparrow \downarrow 
\ldots \uparrow \downarrow\rangle$ is not an eigenstate of the
Heisenberg Hamiltonian. This problem was solved for $g=0$
in the thermodynamic limit $L \rightarrow \infty$ in 1938 by
Hulth\'en \cite{Hulthen} using Bethe's ansatz.

Like the BPS state, the antiferromagnetic vacuum state is of very
high symmetry. It should therefore also be of great interest in
gauge theory. Let us then use the BDS Bethe ansatz 
\eqref{bds},\eqref{u},\eqref{bdseng} and compute the higher-loop
corrections to Hulth\'en's solution. As the computation is done
in the thermodynamic limit the BDS equations are perfectly reliable.
The one-loop solution may be found in many textbooks. It is particularly
well described in the lectures \cite{faddeev}. 
Adapting it to the deformed BDS case is completely straightforward.
We will therefore mostly skip the derivation, referring to \cite{faddeev}
for details, and immediately state the result for the energy of
the antiferromagnetic vacuum:
\begin{equation}
\label{energyintegral}
E(0)=L\, \int_{-\infty}^{\infty} du\, 
\frac{\rho(u)}{u^2+\quarter}
\quad \rightarrow \quad 
E(g)=L\, \int_{-\infty}^{\infty} du\, \rho(u)\,
\left(\frac{i}{x^+(u)}-\frac{i}{x^-(u)}\right)\, ,
\end{equation}
where the auxiliary spectral parameter $x$ \cite{BDS} is given by
\begin{equation}
\label{x}
x(u)=
\frac{u}{2}\left(1+\sqrt{1-\frac{2g^2}{u^2}}\right)\;, \quad {\rm with}\quad
x^{\pm}(u)=x(u\pm\frac{i}{2})\, .
\end{equation}
Here $\rho(u)$ is the thermodynamic density of (magnon) excitations.
It is found from solving the Bethe equations, which turn at 
$L \rightarrow \infty$ into a single non-singular integral equation 
for $\rho(u)$:
\begin{equation}
\label{hultheneq}
-\frac{d p(u)}{d u}=2\,\pi\,\rho(u)+ 2\,\int_{-\infty}^{\infty} du'\, 
\frac{\rho(u')}{(u-u')^2+1}\, ,
\end{equation}
where the derivative of the momentum density is,
with $u_{\pm}=u\pm\frac{i}{2}$, given by
\begin{equation}
\label{dp}
-\frac{d p(u)}{d u}\, \Bigg|_{g=0}=\frac{1}{u^2+\quarter}
\quad \rightarrow \quad 
i \frac{d}{du}\log\frac{x^+(u)}{x^-(u)}=
\frac{i}{\sqrt{u_+^2-2g^2}}
-\frac{i}{\sqrt{u_-^2-2g^2}}\, .
\end{equation}
We notice that the r.h.s.~of Hulth\'en's equation \eqref{hultheneq}
does not depend explicitly on the coupling constant $g$ (since the
S-matrix of the BDS Bethe equation, i.e.~the r.h.s.~of \eqref{bds}
does not look different, in the $u$-variables, from the one
of the Heisenberg model). Furthermore, the kernel of the integral equation
is of difference form and the integration range is infinite. The equation
may therefore immediately solved for $\rho(u)$, for all $g$, 
by Fourier transform:
\begin{equation}
\label{density}
\rho(u)\, \Bigg|_{g=0}=\frac{1}{2 \cosh \pi u}
\quad \rightarrow \quad 
\rho(u)=
\int_0^{\infty}\,\frac{dt}{2 \pi}\,
\frac{\cos\left(t u\right)\,
J_0(\sqrt{2} g t)}{\cosh\left(\frac{t}{2}\right)}\, .
\end{equation}
Plugging this result into the energy expression
\eqref{energyintegral} one finds
\begin{equation}
\label{antiferroenergy}
E(0)=L\,2\,\log 2 
\quad \rightarrow \quad 
E(g)=L\,\frac{4}{\sqrt{2}g}\,
\int_0^{\infty}\,\frac{dt}{t}\,
\frac{J_0(\sqrt{2} g t)\,J_1(\sqrt{2} g t)}{1+e^t}\, , 
\end{equation}
where $J_0(t),J_1(t)$ are standard Bessel functions.

Now, it so turns out that the expressions for $\rho(u)$ in
\eqref{density} and $E(g)$ in 
\eqref{antiferroenergy}  are very famous results in the history
of condensed matter theory. The latter is, up to an overall
minus sign, identical to the ground state energy of the
one-dimensional {\it Hubbard model} at half filling. It was shown
to be integrable and solved by Bethe Ansatz
in 1968 by Lieb and Wu \cite{LW}. Since then a very large literature
on the subject has developed. For some good re- and overviews,
see \cite{books}, \cite{booksbis} . The Hubbard model is not quite
a spin chain, but rather a model of $N_0$ itinerant electrons on
a lattice of length $L$. The electrons are spin-$\half$ particles.
Due to Pauli's principle the possible states at a given lattice 
site are thus four-fold:
(1) no electron, (2) one spin-up electron $\up$, (3) one spin-down 
electron $\dn$,
(4) two electrons of opposite spin $\updn:=\up \dn$. 
Hubbard's Hamiltonian reads, in 
one dimension
\begin{equation}
\label{H}
H_{{\rm Hubbard}}=
-t\, \sum_{i=1}^L \sum_{\sigma=\up,\dn}
\left(c^\dagger_{i,\sigma} c_{i+1,\sigma}+
c^\dagger_{i+1,\sigma} c_{i,\sigma}\right)+
t\,U\, \sum_{i=1}^L 
c^\dagger_{i,\up} c_{i,\up}c^\dagger_{i,\dn} c_{i,\dn}\, .
\end{equation}
The operators $c^\dagger_{i,\sigma}$ and $c_{i,\sigma}$
are canonical Fermi operators satisfying the anticommutation relations
\begin{eqnarray}
\{c_{i,\sigma},c_{j,\tau}\}&=&
\{c^\dagger_{i,\sigma},c^\dagger_{j,\tau}\}=0\, ,
\\
\{c_{i,\sigma},c^\dagger_{j,\tau}\}&=&
\delta_{i j}\,\delta_{\sigma \tau}\, .
\nonumber
\end{eqnarray}
We see that the Hamiltonian consists of two terms, a kinetic 
nearest-neighbor hopping term
with strength $t$, and an ultralocal interaction potential with coupling
constant $U$. Depending on the sign of $U$, it leads to 
on-site attraction or repulsion if two electrons occupy the same site.

Comparing the BDS result \eqref{antiferroenergy} with the result
of Lieb and Wu for the ground state energy of the half-filled band,
where the number of electrons equals the number of lattice sites,
i.e.~$N_0=L$, we see that the two energies coincide {\it exactly} under
the identification
\begin{equation}
\label{identify}
t=-\frac{1}{\sqrt{2}\,g}
\qquad \qquad
U=\frac{\sqrt{2}}{g}\, .
\end{equation}
This leads us to the conjecture that the BDS long-range spin chain, where,
by construction, $g$ is assumed to be small, 
is nothing but the strong coupling limit of the Hubbard model
under the identification \eqref{identify}. 
In the following we will show that this is indeed the case,
even away from the antiferromagnetic ground state. 
In fact, we shall demonstrate that it is {\it exactly true at finite} 
$L$ up to ${\cal O}(g^{2 L})$
where the BDS long-range chain looses its meaning. 
This will, however, require the resolution of certain subtleties 
concerning the boundary conditions of the Hamiltonian 
\eqref{H}. As it stands, it will only properly diagonalize the BDS
chain if the length $L$ is odd. It the length is even, we have
to subject the fermions to an Aharonov-Bohm type magnetic flux $\phi$.
The Hamiltonian in the presence of this flux remains integrable and reads
\begin{equation}
\label{Hflux}
H=
\frac{1}{\sqrt{2}\,g}\, \sum_{i=1}^L \sum_{\sigma=\up,\dn}
\left(e^{i \phi_\sigma}\,c^\dagger_{i,\sigma} c_{i+1,\sigma}+
e^{-i\phi_\sigma}\,c^\dagger_{i+1,\sigma} c_{i,\sigma}\right)-
\frac{1}{g^2}\, \sum_{i=1}^L 
c^\dagger_{i,\up} c_{i,\up}c^\dagger_{i,\dn} c_{i,\dn}\, ,
\end{equation}
where the twist is given by\footnote{
For odd $L$ the twist $\phi_\sigma$ could alternatively be chosen as any 
integer
multiple of $\frac{\pi}{L}$, while for even $L$ any odd-integer
multiple of  $\frac{\pi}{2 L}$ is possible. A compact notation which does 
not distinguish the cases $L$ odd or even is $\phi=\frac{\pi(L+1)}{2L}$.}
\begin{eqnarray}
\label{twist}
&\ & \phi_\sigma=\phi\;, \qquad \sigma=\up,\dn\;, \\ \nonumber
&\ & \phi=0~~{\rm for}~~L={\rm odd} \qquad {\rm and} \qquad 
\phi=\frac{\pi}{2 L}~~{\rm for}~~L={\rm even.}
\end{eqnarray}
An alternative way to introduce the Aharonov-Bohm flux is to 
perform a suitable gauge transformation and to thereby concentrate the 
magnetic potential on a single link, say the one connecting the $L$'th and
the first site.
It is then clear that considering a non-zero flux amounts to considering 
twisted boundary conditions for the fermions.

The vacuum of the Hamiltonian \eqref{H} is the empty lattice of length $L$.
Here the elementary excitations are up ($\up$) and down ($\dn$) spins.
Two electrons per site ($\updn$) are considered a bound state
of elementary excitations. These constituents of the bound states are 
repulsive (as $g>0$).
For our purposes it is perhaps more natural to
consider the BPS vacuum:
\begin{equation}
\label{BPS}
|Z^L\rangle = |\up \up \ldots \up \up \rangle=
c^\dagger_{1\up} c^\dagger_{2\up} \ldots 
c^\dagger_{L-1 \up} c^\dagger_{L \up}\,|0\rangle
\end{equation}
We may then perform a particle-hole transformation on the up-spin electrons. 
\begin{eqnarray}
\label{particlehole}
\hole &\Longleftrightarrow& \up
\\
\dn &\Longleftrightarrow& \updn
\end{eqnarray}
Now single up-spins ($\up$) are considered to be empty sites,
while the elementary excitations are holes ($\hole$) and 
two electrons states ($\updn$). 
In the condensed matter literature, such a transformation
is often called a Shiba transformation
and it is known to reverse the sign of the interaction. The standard 
Shiba transformation contains an alternating sign in the definition
of the new creation/annihilation operators, designed to recover 
the hopping term, at least in the periodic case. The price to pay is 
that for odd lengths  the sign of the
hole hopping term will change on the link connecting the last
($L$'th) and the first site. In other words, the particle/hole transformation
introduces an extra flux of $\pi\, L$ seen by holes.
Since we prefer to distribute this twist 
uniformly along the chain, we remove the signs
in the definition of the hole operators\footnote{This amounts to 
a gauge transformation.} and  put   
\begin{eqnarray}
\label{Shiba}
c_{i,\hole}=c^\dagger_{i,\up}\;, &\ &
\qquad c^\dagger_{i,\hole}=c_{i,\up}\;,
\\
c_{i,\updn}= c_{i,\dn}\;, &\ &
\qquad c^\dagger_{i,\updn}= c^\dagger_{i,\dn}\;.
\end{eqnarray}
Under the particle/hole
transformation, the charge changes sign and the corresponding hopping terms
get complex conjugated. An extra minus sign comes from the reordering of the
hole operators. 
Therefore we may 
write the  Hamiltonian in its dual form 
\begin{equation}
\label{Hdual}
H=
\frac{1}{\sqrt{2}g}\, \sum_{i=1}^L \sum_{\sigma=\hole,\updn}
\left(e^{i\phi_\sigma}\,c^\dagger_{i,\sigma} c_{i+1,\sigma}+
e^{-i\phi_\sigma}\,c^\dagger_{i+1,\sigma} c_{i,\sigma}\right)-
\frac{1}{g^2}\, \sum_{i=1}^L 
(1-c^\dagger_{i,\hole} c_{i,\hole})c^\dagger_{i,\updn} c_{i,\updn}\, .
\end{equation}
where $\phi_\updn=\phi_\dn$, while  $\phi_\hole=\pi-\phi_\up$.
Comparing the two expressions (\ref{Hflux}) and (\ref{Hdual}) we conclude that
under the duality transformation, the  Hamiltonian   (\ref{Hflux})  
transforms as
\begin{equation}
\label{dualrelation}
 H(g;\phi,\phi) \to -H(-g;\pi-\phi,\phi)-\frac{M}{g^2}
\end{equation}
As predicted, the sign of the interaction changes upon dualization. 
The effect is  that holes $\hole$ and states with two electrons per site 
$\updn$ attract each other and form bound states $\dn$, 
the magnons.

\section{Effective Three-Loop Spin Hamiltonian}
\label{sec:effective}

In this chapter we will explicitly demonstrate that the Hubbard
Hamiltonian \eqref{H} generates at small $g$ the three-loop
dilatation operator of $\cN=4$ gauge theory in the $\alg{su}(2)$
sector \cite{BKS}. The BDS long-range spin Hamiltonian \cite{BDS} 
is thus seen to emerge as an {\it effective} Hamiltonian from the 
underlying short-range system. Note that the small $g$ limit,
relevant to perturbative gauge theory, corresponds, via
\eqref{identify}, to the {\it strong coupling} limit 
$U \rightarrow \infty$ of the Hubbard model in condensed matter parlance. 

Our claim may be verified immediately to two-loop order, using 
well-known results in the literature. Klein and Seitz \cite{KS}
proposed the strong-coupling expansion of the 
half-filled Hubbard model to $\cO(g^7)$.
The two-loop result $\cO(g^4)$ was later confirmed by
Takahashi \cite{taka1}. In fact, eq.~(2.15) of his paper\footnote{
Incidentally, this is the famous paper where the next-nearest
neighbor correlation function of the Heisenberg antiferromagnet
was first obtained. We took this as a hint
that the half-filled Hubbard model ``knows'' something about
long-range deformations of the Heisenberg model.}
precisely agrees with the two-loop
piece of the BDS Hamiltonian (and therefore with two-loop
gauge theory \cite{BKS}) under the parameter identification
\eqref{identify}. Eq.~(2.15) of \cite{taka1} also contains
certain four-spin terms which only couple, since our
system is one-dimensional, to a length $L=4$ ring. These
are a first manifestation of certain unwanted terms
which we need to eliminate by appropriate boundary conditions and
twisting, see \eqref{Hflux},\eqref{twist}, to be discussed in more 
detail below. 

When one now turns to the three-loop $\cO(g^6)$ result 
of Klein and Seitz \cite{KS} as obtained in 1972, one unfortunately 
finds that their effective Hamiltonian disagrees with the BDS 
Hamiltonian at this order. We have been unable to find a later
paper in the vast condensed matter literature on the subject
which confirms or corrects their 33 year old calculation.  
We have therefore decided to check their computation in detail.
And indeed we found a mistake, see below. Correcting it, we reproduce 
the planar three-loop $\alg{su}(2)$ dilatation operator
\cite{BKS,dynamic,EJS}, see 
\eqref{hamexpansion},\eqref{hamexpansion2} below.

For the remainder of this chapter 
it is convenient to use (\ref{H}) and rewrite it (see 
Appendix \ref{app:effective3} 
for a discussion about the relevance of the twist factors in computing 
the effective Hamiltonian)  in the form:
\begin{equation} 
\label{hubbard}
H_{{\rm Hubbard}}=
-\, \sum_{i<j}^L t_{i j} (X_{i j} + X_{j i})+
t\,U\, \sum_{i=1}^L 
c^\dagger_{i,\up} c_{i,\up}c^\dagger_{i,\dn} c_{i,\dn}\, .
\end{equation}
where $X_{i j}=\sum_{\sigma=\up,\dn} c^\dagger _{i,\sigma} c_{j, \sigma}$ 
and $t_{i j}=t \ \delta_{i+1,j}$ .

\subsection{Generalities}
\label{sec:generalities}

The Hamiltonian (\ref{hubbard}) consists of two parts: A {\it hopping term} 
involving the coefficients $t_{i j}$, and the {\it atomic part}.
The latter is diagonalized by eigenstates describing localized electrons 
at sites $x_i$. The ground-state subspace 
$\mathcal{E}_0$ of the atomic part is spanned by  
$c^{\dagger}_{1 \tau_1}\, c^{\dagger}_{2 \tau_2} \ldots 
c^{\dagger}_{L-1, \tau_{L-1}}\, c^{\dagger}_{L \tau_L} |0\rangle$.
Here we are interested in the limit of large $U$, with $t$ staying 
relatively small. The atomic part tends to localize the electrons, while 
some hopping may still occur. At low temperatures this corresponds to small
fluctuations around $\mathcal{E}_0$ states, since each hopping of the electron
from one site to another is suppressed by a factor of order of $1/U$. 
One can now pose the question whether it is possible, for large $U$ and low 
temperature, to find an effective operator $h$ acting in $\mathcal{E}_0$ 
whose eigenvalues
\begin{equation} \label{r2}
h |\phi\rangle=E|\phi\rangle
\end{equation}
are the same as for the one of the Hamiltonian (\ref{H}):
\begin{equation} \label{uselessequation}
H |\psi \rangle=E |\psi \rangle\, .
\end{equation}
The answer is to the positive and has a long history \cite{K}. A formal and 
rigorous treatment of this subject is presented in appendix 
\ref{app:effective}. 

It is however instructive to discuss (\ref{r2}) in a more heuristic way. 
It is obvious that the effective Hamiltonian $h$ must properly include the
hopping effects. 
On the other hand it acts only in a subspace of the full state space, where
configurations with double occupancies are projected out. This means that 
(\ref{r2}) should describe processes with {\it virtual} intermediate states, 
corresponding to electrons hopping from site $i$ to site $j$ and subsequently 
hopping back. Since every nearest-neighbor hopping is suppressed by $1/U$ 
it is clear that the $1/U$ expansion of $h$ will result in increasingly 
long-range interactions. What kind of terms may appear in $h$? 
A first guess leads to products of hopping operators $X_{ij}$ with the 
condition that they will not move states out of the space $\mathcal{E}_0$. 
Since $X_{ij}$ annihilates an electron at site $j$ and creates a new one at 
$i$, we see that only such products of $X_{ij}$ operators are allowed 
which result in the same number of creation and annihilation operators at 
a given lattice site. Since each product of creation and annihilation 
operators may be represented in terms of $\alg{su}(2)$ spin operators, 
we conclude that the effective Hamiltonian (\ref{r2}) must be of spin-chain 
form! 

\subsection{Three-Loop Result}

We have used perturbation theory for degenerate systems 
(see appendix \ref{app:effective1}, 
where also some details of the computation scheme are explained) 
to derive the effective Hamiltonian to three loops.
The result up to sixth order (i.e.~three loops) for the formal perturbation
theory expansion is found in \eqref{formal} in appendix
\ref{app:effective2}.
It may be shown to be completely equivalent to the expansion obtained 
by Klein and Seitz in \cite{KS}.

The formal expansion is then converted into a diagrammatic expansion,
see again appendix \ref{app:effective2}. 
We agree with Klein and Seitz with all perturbation theory diagrams up 
to sixth order as presented in their paper, except that we find that they  
missed a few diagrams (of type '$a$' as in Fig 5.~of their paper).
These are the following diagrams (summation over $i$ is understood): 
\vspace{6 mm}
\begin{center}
\includegraphics[scale=0.5]{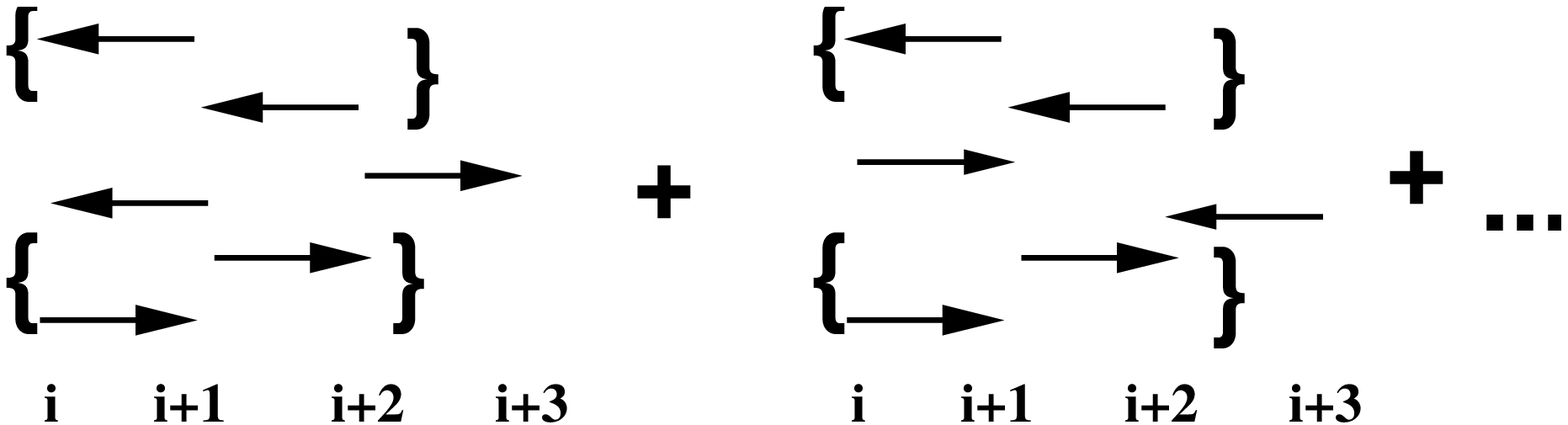}
\end{center}
\vspace{6mm}
where $\ldots$ means arrow-reversed diagrams. 

We have confirmed all diagram evaluations performed in \cite{KS}, 
except for the contribution of the diagrams of type $f$ in equation (C3) 
of the mentioned paper, where there is an overall factor of 16 missing. 
We believe this to be a typographical error. There is however also an 
additional contribution from the mentioned four diagrams which were not 
included in their computations. Explicit calculation shows, that the
missing terms yield
\begin{equation} 
\label{contribution}
-\bigg(\frac{1}{U} \bigg)^6 U t (16 A_1-4 A_2+2 B_3-2 B_1-2 B_2)\, ,
\end{equation}
where
\begin{displaymath}
\begin{array}{c} 
A_s= \sum_{i=1}^L (1-\mathcal{P}_{i,i+s})\, , \qquad 
B_1= \sum_{i=1}^L (1-\mathcal{P}_{i,i+1}\mathcal{P}_{i+2,i+3})\, , \\ \\
B_2=\sum_{i=1}^L (1-\mathcal{P}_{i,i+2}\mathcal{P}_{i+1,i+3})\, , \qquad  
B_3=\sum_{i=1}^L (1-\mathcal{P}_{i,i+3}\mathcal{P}_{i+1,i+2})\, ,
\end{array}
\end{displaymath}
and $\mathcal{P}$ is a spin permutation operator. Correcting the result of
Klein and Seitz we find
\begin{equation} 
\label{main}
\begin{array}{c}h=\bigg[\displaystyle -2 \bigg( \frac{1}{U}\bigg)^2+8 
\bigg(\frac{1}{U}\bigg)^4-56 \bigg(\frac{1}{U}\bigg)^6\bigg]t U A_1 
+\bigg[-2\bigg(\frac{1}{U}\bigg)^4
+16 \bigg(\frac{1}{U}\bigg)^6\bigg] t U A_2 \ 
+ \\ \\ \displaystyle 4 \bigg(\frac{1}{U}\bigg)^6  t U (B_2-B_3)\, .
\end{array}
\end{equation}
Upon putting $U=\frac{\sqrt{2}}{g}$, $t=-\frac{1}{\sqrt{2} g}$ and after 
some simple algebra one rewrites (\ref{main}) in the form
\begin{equation} 
\label{hamexpansion}
h=\sum_{i=1}^L (h_2 +g^2 h_4 +g^4 h_6 +...)\, ,
\end{equation}
with
\begin{eqnarray}
\label{hamexpansion2}
h_2 &=& \frac{1}{2}(1-\vec{\sigma_{i}} \vec{ \sigma_{i+1}})\, , \nonumber \\
h_4 &=&-(1-\vec{\sigma_{i}}\vec{\sigma_{i+1}})
+\frac{1}{4}(1-\vec{\sigma_{i}}\vec{\sigma_{i+2}})\, , \nonumber\\
h_6 &=& \frac{15}{4} (1-\vec{\sigma}_{i} \vec{\sigma}_{i+1})
-\frac{3}{2} (1-\vec{\sigma}_{i}\vec{\sigma}_{i+2})
+\frac{1}{4}(1-\vec{\sigma_{i}}\vec{\sigma}_{i+3}) \nonumber \\
&&-\frac{1}{8}(1-\vec{\sigma_{i}}\vec{\sigma}_{i+3})(1-\vec{\sigma}_{i+1}
\vec{\sigma}_{i+2})\nonumber
\\ && +\frac{1}{8}(1-\vec{\sigma_{i}}\vec{\sigma}_{i+2})(1-\vec{\sigma}_{i+1}
\vec{\sigma}_{i+3})\, .
\end{eqnarray}
This is indeed the correct planar three-loop dilatation operator in the
$\alg{su}(2)$ sector of $\cN=4$ gauge theory \cite{BKS}.
It is fascinating to see its emergence from an important and
well-studied integrable model of condensed matter theory.

\section{Lieb-Wu Equations}
\label{sec:liebwu}

The Hamiltonian \eqref{H} was shown to be integrable and diagonalized
by coordinate Bethe ansatz in \cite{LW}. 
For a pedagogical treatment see \cite{books}. 
This required finding the dispersion relation of the elementary
excitations $\up$ and $\dn$ and working out their two-body S-matrix.
It is indeed a {\it matrix} since there are two types of
excitations, hence their ordering matters. The scattering of
two up- or two down-spins is absent, as identical fermions behave
like free particles. The scattering of different types of fermions
is non-trivial due to their on-site interaction. After working
out the S-matrix one needs to diagonalize the multi-particle system
by a {\it nested} Bethe ansatz. The result of this procedure, 
generalized to the case with magnetic flux, yields 
the Lieb-Wu equations:
\begin{eqnarray}
\label{liebwu1}
&\ &e^{i\tilde q_nL}=\prod_{j=1}^M 
\frac{u_j-\sqrt{2}g\sin (\tilde q_n+\phi) -i/2}
{u_j-\sqrt{2}g\sin  (\tilde q_n+\phi)+i/2}\, ,
\qquad n=1,\ldots, L
\\
\label{liebwu2}
&\ &\prod_{n=1}^{L}
\frac{u_k-\sqrt{2}g\sin  (\tilde q_n+\phi) +i/2}
{u_k-\sqrt{2}g\sin  (\tilde q_n+\phi)- i/2}= 
\prod_{\textstyle\atopfrac{j=1}{j\neq k}}^M
\frac{u_k-u_j +i}{u_k-u_j-i}\, ,
\quad k=1,\ldots, M
\end{eqnarray}
where the twist is given\footnote{
The Lieb-Wu equations for arbitrary twist are given in appendix 
\ref{app:generic}.} 
in \eqref{twist} and the energy is
\begin{equation}
\label{liebwueng}
E=\frac{\sqrt{2}}{g}\;
\sum_{n=1}^{L}\cos  (\tilde q_n+\phi)\;.
\end{equation}
Here we have already specialized to the half-filled case with
$N_0=L$ fermions and $M\leq L/2$ down-spin fermions (there are
thus $L-M$ up-spin fermions in the system).

This form of the Hubbard model's Bethe equations if very convenient
for demonstrating rather quickly that the $g\rightarrow 0$ limit
yields the spectrum of the Heisenberg magnet. In fact, the 
Lieb-Wu equations decouple at leading order and become
\begin{eqnarray}
\label{XXX1}
&\ &e^{i\tilde q_nL}=\prod_{j=1}^M 
\frac{u_j -i/2}
{u_j+i/2}\,\Big(1+{\cal O}(g)\Big)\, ,
\qquad n=1,\ldots, L
\\
\label{XXX2}
&\ &\left(
\frac{u_k +i/2}
{u_k- i/2}\right)^L= 
\prod_{\textstyle\atopfrac{j=1}{j\neq k}}^M
\frac{u_k-u_j +i}{u_k-u_j-i}\, ,
\quad k=1,\ldots, M
\end{eqnarray}
Eqs.~\eqref{XXX2} are already identical to the ones of the
Heisenberg magnet (see e.g.~\cite{MZ},\cite{faddeev}).
The r.h.s.~of \eqref{XXX1} is, to leading order ${\cal O}(g^0)$, 
the eigenvalue of the shift operator
of the chain (again, \cite{MZ},\cite{faddeev}). In gauge theory
we project onto cyclic states, so we may take the eigenvalue to be one, and 
solve immediately for the $L$ momenta $\tilde q_n$ to leading order:
\begin{equation}
\label{dualmom}
e^{i\tilde q_nL}=1
\qquad \Longrightarrow \qquad
\tilde q_n=\frac{2 \pi}{L} (n-1) + {\cal O}(g), \quad n=1,\ldots, L\, .
\end{equation}
But now we have to find the energy. Plugging the result \eqref{dualmom}
into the expression \eqref{liebwueng} conveniently 
eliminates the spurious ${\cal O}(1/g)$ term in the energy.
We therefore need to find the ${\cal O}(g)$ corrections to the
momenta in \eqref{dualmom} from \eqref{liebwu1}.
Luckily, this is a {\it linear} problem; solving it one computes
the ${\cal O}(g^0)$ term of \eqref{liebwueng} as
\begin{equation}
\label{XXXeng}
E=\sum_{k=1}^M\,\frac{1}{u_k^2+\frac{1}{4}}\,+{\cal O}(g)\, ,
\end{equation}
which is the correct expression for the energy of the 
Heisenberg magnet.

The starting point for the small $g$ expansion of the
Lieb-Wu equations are therefore Bethe's original equations
\eqref{XXX2},\eqref{XXXeng} in conjunction with the free particle
momentum condition \eqref{dualmom}. It is interesting that all
non-linearities are residing in the one-loop Bethe equations \eqref{XXX2}.
Once these are solved for a given state, the perturbative 
expansion is obtained from a  linear, recursive procedure.
It allows for efficient and fast numerical computation of
the loop corrections to any state once the one-loop solution is known.
A simple tool for doing this with e.g.~\texttt{Mathematica} may
be found in Appendix \ref{app:code}, along with a similar tool for the
perturbative evaluation of the BDS equations.

We have applied this perturbative procedure to all\footnote{
The only exception are certain singular three-magnon states which 
require a special treatment.} 
(cyclic) states of the BDS chain
as recorded, up to five loops, in Table 1, p.30 of \cite{BDS}. 
The (twisted) Lieb-Wu equations 
\eqref{liebwu1},\eqref{liebwu2},\eqref{liebwueng} perfectly reproduce
the energies of this table.

We found that that our version of the Hubbard model {\it precisely
agrees} in all investigated cases with the results of the BDS ansatz
up to and including the $(L-1)$-th loop order. On the other
hand, invariably, at and beyond $L$'th order of perturbation theory
(corresponding to the ${\cal O}( g^{2L-2})$ terms in the energy $E$)
the predictions of the two ans\"atze {\it differ}.
See chapter \ref{sec:konishi} for some concrete examples.

It is also interesting to record the effects of the twists on 
the perturbative spectrum. A first guess might be that they
should only influence the spectrum at and beyond wrapping order 
${\cal O}( g^{2L-2})$, when the order of the effective
interactions reaches the size of the ring, and the system should become
sensitive to the boundary conditions. In actual fact, however,
one finds that the twists generically influence the spectrum starting at
already ${\cal O}( g^{L-2})$. This is the phenomenon of 
{\it demi-wrappings}. The Hubbard model at small $g$ behaves effectively
as a long-range spin chain due to the virtual ``off-shell'' decomposition 
of the magnon bound states $\dn$ (which are sites occupied by a down-spin 
but no
up-spin) into holes $\hole$ (empty sites) and double-occupied sites $\updn$. 
The power of the coupling constant $g$ counts the number of steps 
a hole $\hole$ or $\updn$-particle is exercising during its virtual
excursion, see also the discussion in section \ref{sec:generalities}. 
We now observe that starting from at ${\cal O}( g^{L-2})$ the 
excitations $\hole$,$\updn$ can (virtually) travel around the ring,
and the amplitudes start to depend on the boundary conditions!
A similar distinction between wrappings and demi-wrappings was 
qualitatively discussed in a recent paper on this subject 
\cite{AJK}. Our procedure of twisting eliminates the demi-wrappings.
Interestingly, this seems to leave no further freedom at and beyond
wrapping order, at least in the context of our current
construction.

The Lieb-Wu equations in the form 
\eqref{liebwu1},\eqref{liebwu2},\eqref{liebwueng} are very useful
for the analysis of chains of small length. They are far less convenient
in or near thermodynamic situations, i.e.~when $L \rightarrow \infty$. The
reason is the large number of momenta $\tilde q_n$ one has to deal with.
In \eqref{Hdual} we have written a dual form of the Hamiltonian
\eqref{Hflux}. Accordingly, we may write down the
corresponding set of {\it dual} Lieb-Wu equations:
\begin{eqnarray}
\label{liebwudual1}
&\ &e^{i q_nL}=\prod_{j=1}^M 
\frac{u_j-\sqrt{2}g\sin (q_n-\phi) -i/2}
{u_j-\sqrt{2}g\sin  (q_n-\phi)+i/2}\, ,
\qquad n=1,\ldots, 2 M
\\
\label{liebwudual2}
&\ &\prod_{n=1}^{2M}
\frac{u_k-\sqrt{2}g\sin  ( q_n-\phi) +i/2}
{u_k-\sqrt{2}g\sin  (q_n-\phi)- i/2}=
-\prod_{\textstyle\atopfrac{j=1}{j\neq k}}^M
\frac{u_k-u_j +i}{u_k-u_j-i}\, ,
\quad k=1,\ldots, M
\end{eqnarray}
where the energy is now given by
\begin{equation}
\label{liebwudualeng}
E=-\frac{M}{g^2}-\frac{\sqrt{2}}{g}\;
\sum_{n=1}^{2M}\cos  (q_n-\phi)\;.
\end{equation}
Again, we have specialized to the case of half-filling. A particular feature 
of the 
dual Hamiltonian \eqref{Hdual} is that the twist is different for the 
two components.
We are therefore led to use the
Lieb-Wu equations for generic twist  which are written down in 
Appendix \ref{app:generic}.
This explains the minus sign in the right hand of \eqref{liebwudual2}, 
$e^{i(\phi_\updn-\phi_\hole)}
=e^{i(2\phi-\pi)L}=-1$.
Note that $\phi\to -\phi$ is a symmetry of the equations (but not of the
solutions), as we may change
$u\to-u$ and $q\to-q$.
Note also that therefore the set of $L+2 M$ momenta 
$(\tilde q_n,-q_n)$ corresponds to
the $L+2 M$ solutions of the first Lieb-Wu equation \eqref{liebwu1}.

\section{Magnons from Fermions}
\label{sec:bdsproof}

In chapter \ref{sec:effective} we proved, to three-loop order, 
that the Hamiltonian of the BDS long-range spin chain emerges
at weak coupling $g$ from the twisted Hubbard Hamiltonian
as an effective theory. Pushing this proof to higher orders would
be possible but rather tedious. Note, however, that the BDS
Hamiltonian is, at any rate, only known to five-loop order 
\cite{BDS}. What we are really interested in is whether the
Bethe ansatz \eqref{bds},\eqref{u},\eqref{bdseng}, which was
{\it conjectured} in \cite{BDS}, may be {\it derived} from the
Bethe equations of the Hubbard model, i.e.~from the 
Lieb-Wu equations of the previous chapter. We will now show 
that this is indeed the case. The derivation will first focus
on a single magnon (section \ref{sec:bdsproof1}), where it will
be shown that the magnons $\dn$ of the long-range spin chain
are bound states of holes $\hole$ and double-occupations $\updn$,
as is already suggested by the perturbative picture of
chapter \ref{sec:effective}. It will culminate in 
\ref{sec:bdsproof3}, where we demonstrate that the bound states
alias magnons indeed scatter according to the r.h.s.~of
\eqref{bds}. An alternative proof may be found in appendix 
\ref{app:alternative}.

Unlike the BDS long-range spin chain, the twisted Hubbard model
is well-defined away from weak coupling, and actually
for arbitrary values of $g$.
An important question is whether the twisted Hubbard model
allows to explain the vexing discrepancies between gauge and string
theory \cite{Call1,SS}. Unfortunately this does not seem to be
the case. We have carefully studied the spectrum of
two magnons in section \ref{sec:bdsproof2}, and find that
their is no order of limits problem as the coupling $g$ and the
length $L$ tend to infinity while $g/L$ stays finite.
The scattering phase shift indeed always equals the one
predicted by the BDS chain.

The Hubbard model contains also many states which are separated
at weak coupling by a large {\it negative} energy gap 
$\cO(-1/g^2)$ from the magnons. This may be seen from the
expression \eqref{liebwudualeng}. For solutions with real momenta
$q_n$ the cosine is bounded in magnitude by one, and the 
constant part $-M/g^2$ cannot be compensated. These states are
composed of, or contain, holes $\hole$ and double-occupations $\updn$
which are unconfined, i.e.~which do not form bound states.
Their meaning will need to be understood if it turns out that the 
$\cN=4$ gauge theory's dilatation operator can indeed be described 
by a Hubbard model beyond the perturbative three-loop approximation.
In fact, it is clear from the expression for the
anomalous dimension $\Delta=L+g^2\, E(g)$ in \eqref{ops} 
that each unconfined pair $(\hole,\updn)$
shifts the classical dimension and thus the length down by one:
$L \rightarrow L-1$. Is this a first hint that the perturbative
$\alg{su}(2)$ sector of $\cN=4$ gauge theory does not stay
closed at strong coupling, as was argued in \cite{Joe}?

\subsection{One-Magnon Problem}
\label{sec:bdsproof1}

Let us then begin by studying the case of $M=1$ down spin and $L-1$
up spins, see \cite{books},\cite{EsslerFinite}. 
Clearly it is easiest to use the dual form of the Lieb-Wu
equations \eqref{liebwudual1},\eqref{liebwudual2},\eqref{liebwudualeng}.
In the weakly coupled spin chain we have only $L$ states, while
in the Hubbard model we have $L^2$ states. This is because
one down spin $\dn$ is composed of one hole $\hole$ and 
one double-occupation $\updn$. If we project to cyclic states, as
in gauge theory, only one of the $L$ states survives, namely the
zero-energy BPS state. However, in order to derive the magnon dispersion
law, we will not employ the projection for the moment. This
way the magnon can carry non-zero momentum and energy. 
In the Hubbard model the magnon should be a $\hole -\!\!\updn$ bound state,
and we therefore make the ansatz (with $\beta>0$ and $q>0$):
\begin{equation}
\label{1ansatz1}
q_1-\phi=\frac{\pi}{2}+q+i\,\beta\, ,
\qquad
q_2-\phi=\frac{\pi}{2}+q-i\,\beta\, .
\end{equation}
Here $q_1$ and $q_2$ are the quasimomenta of the $\hole$ and the $\updn$
particles. They are complex, where the imaginary part $\beta$ describes
the binding. Adding the real parts gives the momentum $2 q$ of the
magnon. The dual Lieb-Wu equations for one magnon, where we
only have a single rapidity $u$, read
\begin{eqnarray}
\label{1magnon1}
&\ &e^{i q_1L}= 
\frac{u-\sqrt{2}g\sin (q_1-\phi) -i/2}
{u-\sqrt{2}g\sin  (q_1-\phi)+i/2}\, ,
\quad
e^{i q_2L}= 
\frac{u-\sqrt{2}g\sin (q_2-\phi) -i/2}
{u-\sqrt{2}g\sin  (q_2-\phi)+i/2}\, ,
\\
\label{1magnon2}
&\ &
\frac{u-\sqrt{2}g\sin  ( q_1-\phi) +i/2}
{u-\sqrt{2}g\sin  (q_1-\phi)- i/2}\,
\frac{u-\sqrt{2}g\sin  ( q_2-\phi) +i/2}
{u-\sqrt{2}g\sin  (q_2-\phi)- i/2}=
-1\, .
\end{eqnarray}
By multiplying, respectively, the left and right sides of the two 
equations in \eqref{1magnon1} and using 
\eqref{1ansatz1},\eqref{1magnon2} we derive
\begin{equation}
\label{modes}
e^{i\, 2 q\, L}=1
\qquad \Rightarrow \qquad
q=\frac{\pi}{L}\, n\quad (n=0,1,\ldots,L-1)\, .
\end{equation}
This is just the statement that the magnon is free (there is nothing
to scatter from)  and its
momentum $p:=2 q$ is quantized on the ring of length $L$.
Furthermore we can rewrite \eqref{1magnon1} as 
\begin{equation}
\label{cotform}
\sqrt{2}\,g\,\sin \left(q_{1,2}-\phi\right)-u=
\frac{1}{2}\,\cot\left(\frac{q_{1,2}\, L}{2}\right)\, .
\end{equation}
Decomposing into real and imaginary parts we find, using
the twist \eqref{twist},
\begin{equation}
\label{sinhbeta}
\sinh\left(\beta\right)=
\frac{1}{2\sqrt{2}g}\,\frac{1}{\sin\left(q\right)}\,
\tanh\left(\beta\,L\right)\, ,
\end{equation}
and\footnote{
The sign of the second term in \eqref{urelation} may be changed by 
choosing a different gauge for the twist. This type of gauge dependence 
should not appear in physical observables such as the energy.
}
\begin{equation}
\label{urelation}
u=\sqrt{2}g\,\cos\left(q\right)\,\cosh\left(\beta\right)+
\frac{(-1)^n\,(-1)^{\frac{L+1}{2}}}{2\,\cosh\left(\beta\,L\right)}\, .
\end{equation}

By analyzing \eqref{sinhbeta} we may now discuss the existence
of bound states. We see that for large $L$, where 
$\tanh\left(\beta L\right)\rightarrow 1$, we have, for given mode number $n$, 
exactly one\footnote{
Actually, if $g$ becomes of the order of $L$ such that $g/L$ is larger
than a certain threshold value, the bound state is lost. An additional
real solution, c.f.~appendix \ref{app:one-magnon}, will appear.
}
solution with
$\beta>0$ for {\it all values of} $g>0$. We also see that
there is only one way to take the thermodynamic limit, independent of $g$:
\begin{equation}
\label{sinhbeta2}
\sinh\left(\beta\right)\simeq
\frac{1}{2\sqrt{2} \pi}\,\frac{1}{n g/L}\, .
\end{equation}
But this means that there is also only one way to take the
BMN scaling limit, where $g,L \rightarrow \infty$ with $g/L$ kept finite.

Let us then work out the energy of the magnon with momentum $p=2 q$.
The exponential terms $\tanh(\beta L) \simeq 1-2 e^{-2 \beta L}$ 
may clearly be neglected at large $L$ for arbitrary values of $g$, and
we immediately find the dispersion law 
\begin{equation}
\label{dispersion}
E=-\frac{1}{g^2}+\frac{2 \sqrt{2}}{g}\,\sin\left(\frac{p}{2}\right)\,
\cosh (\beta)=
-\frac{1}{g^2}+\frac{1}{g^2}
\sqrt{1+8g^2\sin^2\frac{p}{2}}\, ,
\end{equation}
which is exactly the BDS result \eqref{bdseng}!
Likewise, again dropping the exponential terms from the 
rapidity relation \eqref{urelation}, we find the 
BDS result \eqref{u} for the dependence of the rapidity $u(p)$
on the momentum $p_k=p=2 q$.

Note that our derivation only assumed the thermodynamic limit;
it did {\it not} assume weak coupling. If the coupling $g$ is weak
we may in addition deduce from \eqref{sinhbeta} that the binding 
amplitude $\beta$
diverges logarithmically as $\beta \simeq -\log g$. We may then
deduce that the exponential terms we dropped are
\begin{equation}
\label{exponential}
e^{-2 \beta\, L} \simeq g^{2 L}\, ,
\end{equation}
and therefore should be interpreted as
${\cal O}(g^{2 L})$ wrapping corrections.

We just showed that $L$ of the $L^2$ states of the Hubbard model's
$M=1$ states can be interpreted as magnons. The remaining $L(L-1)$
states should correspond to solutions where
the momenta $q_1,q_2$ are {\it real}, i.e.~these are not bound states.
Among these, $L-1$ states are cyclic. The unbound states are found as
follows. We make the ansatz
\begin{equation}
\label{1ansatz2}
q_1-\phi=\frac{\pi}{2}+q+b\, ,
\qquad
q_2-\phi=\frac{\pi}{2}+ q-b\, ,
\end{equation}
which is completely general except for the
assumption that $q$ and $b$ are {\it real}.
The twisted dual Lieb-Wu equations \eqref{1magnon1},\eqref{1magnon2}
still apply, and, using the same multiplication trick as before 
we find again \eqref{modes}.
%
%
This is merely the statement that 
that the total momentum is quantized on the ring of length $L$.
The Lieb-Wu equations \eqref{1magnon1} now read
\begin{equation}
\label{freecotform}
\sqrt{2}\,g\,\cos \left(q\pm b\right)-u=
\frac{1}{2}\,
\cot\left((q \pm b+\frac{\pi}{2}+\phi)\,\frac{ L}{2}\right)\, .
\end{equation}

Let us first consider the unconfined {\it cyclic} states,
i.e.~the case of mode number $n=0$ in \eqref{modes}, 
hence $q=0$. We can then immediately
find the energy of such states to-be from \eqref{liebwudualeng}
\begin{equation}
\label{chainsaw}
E=-\frac{1}{g^2}\, ,
\end{equation}
which is seen to not depend on $b$.
But can we really find values for $b$ which satisfy
the Lieb-Wu equations \eqref{freecotform}? 
How many solutions of this type do we have? The answer is easily found from
subtracting either side of the two equations in \eqref{freecotform}.
This yields the consistency condition
\begin{equation}
\frac{1}{2}\,
\cot\left((\frac{\pi}{2}+\phi+b)\,\frac{ L}{2}\right)=
\frac{1}{2}\,
\cot\left((\frac{\pi}{2}+\phi-b)\,\frac{ L}{2}\right)\, .
\end{equation}
Now it is very easy to show that there are {\it precisely}
$L-1$ solutions of this equation:
\begin{equation}
\label{bmodes}
b=\frac{\pi}{L}\, m\, 
\qquad {\rm with} \qquad
m=1,\ldots,L-1\, .
\end{equation}
Therefore, the $M=1$ cyclic unconfined ($L-1$-fold degenerate)
states resemble zero-energy ``BPS states'' with exact
scaling dimension $\Delta=L-1$. However, see appendix
\ref{app:one-magnon}.

Finally, let us study the number of states and
the dispersion law of the unconfined states
carrying non-zero total momentum $p=2 q=\frac{2 \pi}{L}\, n$, 
cf.~\eqref{modes}.
We find that by eliminating $u$ from \eqref{freecotform} that
\begin{equation}
\label{sinbeta}
\sin\left(b\right)=
\frac{1}{2\sqrt{2}g}\,\frac{1}{\sin\left(q\right)}\,
\tan\left(b\,L\right)\, ,
\end{equation}
which turns out to just be the analytic continuation of
\eqref{sinhbeta}. 
It is not hard to prove that there are indeed generically
$L-1$ solutions for each value of the $L-1$ non-zero values of $q$.
This yields $(L-1)^2$ states. Therefore, adding these to 
the $L-1$ cyclic real solutions, and the $L$ bound states, 
we have accounted for all of the $L^2$ states of the $M=1$ problem.
In appendix \ref{app:one-magnon}
we investigate the energy of the real solutions in the large $g$ and 
large $L$ limit. In the limit $L\to \infty$, the solutions of \eqref{sinbeta} 
become dense
on the interval $(0,2\pi)$, so for any value of the magnon momentum $p=2q$
we have a continuum of states, whose energies vary continuously.
It is not clear how one would interpret these states in the context of the 
gauge theory, or, more generally, the AdS/CFT 
correspondence. 
It is possible that we need a model encompassing all the sectors of the 
gauge theory to be able to draw some conclusion
about the large $g$ limit.

Let us now turn to the mutual scattering of our magnons; first for
two, and then for arbitrarily many. We shall find that the scattering is,
up to exponential terms, indeed given by the r.h.s.~of \eqref{bds}.

\subsection{Two-Magnon Problem}
\label{sec:bdsproof2}

The result of the previous section does not bode well for
the hope expressed in the last chapter of \cite{BDS} that
wrapping might explain the discrepancies between gauge
and string theory. This would require an order of limits
problem as one takes the coupling $g$ and the length $L$ large.
It is certainly not seen on the level of bound state formation,
recall \eqref{sinhbeta}. However, one might still hope that
the magnons constructed in the last section might somehow
scatter in distinct ways at weak and strong coupling.
By considering the $M=2$ two-magnon problem we will now show that,
unfortunately, this is not the case.
It therefore seems that the AFS string Bethe ansatz \cite{AFS}
cannot be obtained from the twisted Hubbard model, at least
not in the current version. 

We proceed much as before, making the 
appropriate ansatz for two holes and two double-occupancies 
(with $\beta>0$ and $q>0$) bound into two magnons with momenta
$p=2 q$ and $-p=-2 q$:
\begin{eqnarray}
q_1-\phi&=&\frac{\pi}{2}+q+i\,\beta\, , 
\qquad
q_3-\phi=\frac{\pi}{2}+q-i\,\beta\, ,
\\
q_2-\phi&=&-\frac{\pi}{2}-q+i\,\beta\, ,
\qquad
q_4-\phi=-\frac{\pi}{2}-q-i\,\beta\, .
\nonumber
\end{eqnarray}
We derive (for simplicity assume $L \equiv 1~{\rm mod}~4$, which
allows to assume that the two rapidities obey $u_1=-u_2$) from 
the dual Lieb-Wu equations
\begin{equation}
\label{2msinh}
\sinh\left(\beta\right)=
\frac{\cosh\left(\beta\right)\,\cot\left(q\right)\,\sinh\left(\beta\,L\right)}
{2\sqrt{2}g\,\cos\left(q\right)\cosh\left(\beta\right)
\left(\cosh\left(\beta\,L\right)+\sin\left(qL\right)\right)
-\cos\left(q\,L\right)}\, .
\end{equation}
While looking superficially different, this agrees precisely,
up to exponential terms, in the  $L\rightarrow\infty$ limit with
\eqref{sinhbeta}. It is crucial to note that, as in the previous $M=1$ case,
there is only {\it one} way to take the BMN scaling limit, which
yields again \eqref{sinhbeta2}.
Likewise, we find, up to exponential corrections, for the rapidity
\begin{equation}
\label{2mu}
u^2=\frac{1}{4}+g^2+g^2\,\cos\left(2q\right)\,\cosh\left(2\beta\right)=
-\frac{1}{4}
+g^2(\cos\left(2q\right)+\cosh\left(2\beta\right))\, .
\end{equation}
Unfortunately one may now derive from \eqref{2msinh} and \eqref{2mu} that 
the phase shift when the two magnons scatter at large $L$ is always as in 
the BDS chain, and thus as in the gauge theory's near-BMN limit.  
The exponential corrections disappear in and near the BMN limit.

\subsection{Many Magnons and a Proof of the BDS Formula}
\label{sec:bdsproof3}

We have seen above that magnons $\dn$ can arise as
bound states of holes $\hole$ and doubly occupied sites $\updn$. 
The solutions associated to these bound states are known in the condensed 
matter literature as $k\!-\!\Lambda$ strings\footnote{In our notation, they 
should be called $q\!-\!u$ strings.} and were 
first considered by Takahashi\footnote{
In the repulsive case 
considered by Takahashi, the energy of such a bound state is
greater that the energy of its constituents, 
but the wave function is localized in space, 
so they can still be called bound states.}  
in \cite{TakaStrings}.
The explicit solution of the one and two magnon problem allowed us to  
understand that
the deviation from the ``ideal string'' configuration vanishes exponentially 
with the chain length.
In other terms, the string solutions are asymptotic.  

In this chapter we consider the case of solutions with an arbitrary number 
of magnons.
We are able to show that, in the asymptotic regime $L\to \infty$, 
the scattering of magnons
associated to the bound states discussed above is described by the BDS ansatz.

The finite size corrections may be evaluated, similarly to the one-magnon 
case, to be of the order
$e^{-2\beta L}$ where $\beta$ is the typical strength of the binding $\sinh 
\beta \sim 1/g$.
At weak coupling, or in the perturbative regime, these corrections are 
of order $g^{2L}$, as expected.

For reasons of simplicity, 
we are concentrating first on magnons with real momentum,
that is strings containing only one $u$.
In this situation, the momenta $q_n$ appear in complex conjugate pairs. 
Let us choose the labels such that the first $M$ momenta have a positive 
imaginary part $\beta_n$, while the last $M$ momenta
have a negative imaginary part. With the experience gained from the 
one- and two-magnon case
we denote
\begin{eqnarray}
 &\ & q_n- \phi =s_n\,\frac{\pi}{2}+\frac{p_n}{2}+i \beta_n\;, \\
&\ &  q_{n+M}- \phi =s_n\,\frac{\pi}{2}+\frac{p_n}{2}-i\beta_n\;, 
\quad \beta_n>0\;, \quad n=1,\ldots ,M\;. \nonumber
 \end{eqnarray}
where $p_n$ will be the magnon momentum, and $s_n={\rm sign}\ p_n$\footnote{
We assume that $p_n\in (-\pi,\pi)$, meaning that the real part of $q_n- 
\phi$ 
ranges from $\pi/2$ to $3\pi/2$. 
It is interesting to note that there is 
no consistent solution with $q_n- \phi\in (-\pi/2, \pi/2)$. Such a solution 
would imply a negative energy for the corresponding magnon, which is 
unphysical.}.
If $L$ is large, the left hand side of \eqref{liebwudual1} vanishes 
exponentially for 
$n=1,\ldots ,M$ and diverges
 for $n=M+1,\ldots ,2M$. Therefore, for $L$ infinite and 
for any $n=1,\ldots ,M$  
 there exist one $u$, which will be called $u_n$, such that 
\begin{eqnarray}
 u_n-i/2 =\sqrt{2}g \sin ( q_n -\phi) \;, \qquad  u_n+i/2 =\sqrt{2}g\, 
\sin (  q_{n+M} -\phi)\; ,
 \end{eqnarray} 
 or, equivalently,
 \begin{equation}
 \label{upmbeta}
u_n\pm i/2=\sqrt{2} g\; s_n \cos ( \frac{p_n}{2} \mp i\beta_n)\;.
 \end{equation}
 In particular, equation \eqref{upmbeta} allows to determine the inverse 
size of 
 the bound state, $\beta_n$, in terms of the magnon momentum
 $p_n$
 \begin{equation}
 \label{betapn}
  \sinh \beta_n=\frac{1}{2\sqrt{2}g\; s_n \sin \frac{p_n}{2}}=
\frac{1}{2\sqrt{2}g\; | \sin \frac{p_n}{2}|}\;
 \end{equation}
 and to find the relation between $u_n$ and $p_n$
  \begin{equation}
  u_n=\sqrt{2}g\; s_n\, \cos \frac{p_n}{2}\,\cosh \beta_n=
\frac{1}{2}\cot \frac{p_n}{2}\; \sqrt{1+8g^2\sin^2 \frac{p_n}{2}}\;,
 \end{equation}
 which is nothing else that the relation \eqref{u} of the BDS Bethe ansatz.
 The next step is to eliminate the fermion momenta $q_n$  from the dual 
Lieb-Wu equations and  replace them  by the magnon rapidities $u_n$. In 
order to perform this task, we multiply
 the equations number $n$ and $n+M$ in \eqref{liebwudual1}, so that the 
real parts in the exponential
 mutually cancel 
 \begin{equation}
 \label{bdsbis}
 e^{i(p_n+2 \phi+s_n\pi)L}=
-\prod_{\textstyle\atopfrac{j=1}{j\neq n}}^M\frac{u_n-u_j+i}{u_n-u_j-i}\;.
 \end{equation}
 Under the condition $e^{2i \phi L}=(-1)^{L+1}$, which is satisfied due 
to our choice of the twist
 \eqref{twist},  equation \eqref{bdsbis} is identical to the BDS Bethe 
ansatz equation \eqref{bds}.
The second dual Lieb-Wu equation \eqref{liebwudual2} is automatically 
satisfied, while the energy becomes
\begin{eqnarray}
E&=&-\frac{\sqrt{2}}{g} \sum_{n=1}^M\left(\cos(q_n- \phi)+ 
\cos(q_{n+M}- \phi)\right)-\frac{M}{g^2}\\ \nonumber 
&=&\frac{2\sqrt{2}}{g}\sum_{n=1}^M\left |
\sin\frac{p_n}{2} \right | 
\cosh \beta_n-\frac{M}{g^2}
=\sum_{n=1}^M \frac{1}{g^2}\left(\sqrt{1+8g^2\sin^2\frac{p_n}{2}}-1\right)\;.
\end{eqnarray}
which is, again, the BDS result \eqref{bdseng}.

This proof can be easily extended to the situation when the magnon momenta 
$p_n$ are not all
real. This may be the case for strings containing more than a single $u$. 
We can think of such a string
as being composed of several one- magnon strings, each centered to a 
complex momentum 
$p_n$. The above equations are still valid, under the provision that
$s_n$ is defined as the sign of the real part  of $p_n$, $s_n=
{\rm sign\ Re\;} p_n$. Let us note that
$s_n$ is well defined if $u_n$ is finite.
Of course, $\beta_n$ are not real any more but they are defined by the 
first equality in \eqref{betapn}.

\section{Four-Loop Konishi and the Wrapping Problem}
\label{sec:konishi}

The Hubbard model is capable of naturally dealing with the ``wrapping problem''
\cite{BDS}. The latter is a fundamental difficulty for a long-range
spin chain, where one has to decide how to interpret the Hamiltonian
when the interaction range reaches the size of the system\footnote{
If there are only two-body long-range interactions, as 
e.g.~in the Inozemtsev long-range spin chain \cite{Inoz}, the problem may be
circumvented by periodizing the two-body interaction potential.
If there are also multi-body interactions, as occurs in the 
long-range spin chains appearing in perturbative gauge theory,
it is just not clear how to deal with this problem in a natural fashion.
See \cite{BK} for a very recent discussion of these problems.}.

Let us state the prediction of the Hubbard model for the
anomalous dimension of the lowest non-trivial state, the 
Konishi field, with $L=4$ and $M=2$, to e.g.~eight-loop order\footnote{
It is interesting that the coefficients seem to be all integer, at 
least to the order we checked.}. It is easily obtained using
e.g.~the tool in Appendix \ref{app:code}:
\begin{equation}
E_{{\rm Hubbard}}=6 - 12 g^2 + 42 g^4 - 318 g^6+ 4524 g^8- 63786 g^{10}+
783924 g^{12}- 8728086 g^{14} + \ldots
\end{equation}
The four-loop prediction, $-318\, g^6$, is the first order where wrapping 
occurs. The result should be contrasted to the BDS Bethe ansatz,
which, when we ``illegally'' apply it beyond wrapping order,
yields (again, we used the program described in Appendix \ref{app:code}) 
\begin{equation}
E_{\rm{BDS}}=6 - 12 g^2 + 
    42 g^4 - \frac{705}{4} g^6 + \frac{6627}{8} g^8 - \frac{67287}{16} g^{10} 
+ \frac{359655}{16} g^{12} - \frac{7964283}{64} g^{14}
+\ldots
\end{equation}
We can now see explicitly that the perturbative results for the
energy differs in the two ans\"atze at $\cO(g^6)$, i.e.~four loop 
order.  
The exact result for Konishi is given by a rather intricate 
algebraic curve. Note that the two rapidities $u_1$,$u_2$ are 
{\it not} related by the symmetry $u_1=-u_2$.

Let us likewise contrast the results for the lowest non-BPS state 
with an odd length, namely $L=5, M=2$. The Hubbard model gives
\begin{equation}
E_{{\rm Hubbard}}=4 - 6 g^2 + 17 g^4 - \frac{115}{2} g^6 +
\frac{833}{4} g^8 - \frac{6147}{8} g^{10}  + \frac{44561}{16} g^{12} 
- \frac{303667}{32} g^{14}
\end{equation}
while the BDS ansatz yields
\begin{equation}
E_{\rm{BDS}}=4 - 6 g^2 + 
    17 g^4 - \frac{115}{2} g^6 + \frac{849}{4} g^8 - 
\frac{6627}{8} g^{10} + \frac{53857}{16} g^{12} - \frac{451507}{32} g^{14}
+\ldots
\end{equation}
In line with expectation 
this confirms that the perturbative results for the
energy differ between Hubbard and BDS at $\cO(g^8)$, i.e.~five loop 
order. This is precisely where wrapping first occurs for a length five ring.
The exact result is again given by an intricate algebraic curve.

\section{Conclusions}

The main result of this paper is the identification of
the long-range BDS spin chain of \cite{BDS} as an asymptotic
approximation to a short-range model of itinerant fermions,
the Hubbard model. The latter yields a rigorous microscopic
definition of the former. It furthermore provides the Hamiltonian, which 
was only known, in an ``effective'' form, to five-loop order \cite{BDS}. 
We have explicitly derived the emergence of this effective 
description to three-loop order by correcting a previously
performed strong-coupling expansion of the one-dimensional Hubbard model
\cite{KS}. This establishes and proves that the planar three-loop 
dilatation operator of $\cN=4$ gauge theory is, in the $\alg{su}(2)$ sector, 
generated by a twisted Hubbard model. We have also derived the asymptotic 
Bethe equations of the BDS chain from the Lieb-Wu equations of the 
Hubbard model. 

Our identification allows to resolve the wrapping problem of the 
BDS chain in a, as far as we can currently see, unique fashion. 
It also gives a rigorous definition of 
{\it integrability} beyond wrapping order and therefore 
for a system of {\it finite} extent. 
Recall that the notion of ``perturbative'' integrability
implemented in \cite{BDS} requires, strictly speaking,
an infinite system. This renders the BDS ansatz 
\eqref{bds},\eqref{u},\eqref{bdseng} only asymptotically and
thus approximately valid.
The, admittedly more complicated, Lieb-Wu equations 
\eqref{liebwu1},\eqref{liebwu2},\eqref{liebwueng} or
\eqref{liebwudual1},\eqref{liebwudual2},\eqref{liebwudualeng}
are the generalization of the BDS equations to strictly finite
systems and to arbitrary values of the coupling constant $g$.
Their firm base is an underlying S-matrix satisfying the Yang-Baxter
equation \cite{LW}. What is more, the Hubbard model may be
included into the rigorous framework of the quantum inverse
scattering method. In fact, Shastry discovered its R-matrix
\cite{Shastry}, and Ramos and Martins \cite{RM} diagonalized
the model by algebraic Bethe ansatz. 
These results therefore also embed the BDS spin chain into the
systematic inverse scattering formalism.

We have not been able to find the ``effective'' ansatz
\eqref{bds},\eqref{u},\eqref{bdseng}, which significantly 
simplifies the nested Lieb-Wu equations at half-filling up to wrapping 
terms, in the (vast) literature on the Hubbard model
\cite{books}. 
This striking simplification seems to be a 
discrete and generic generalization of the decoupling phenomenon 
of the system of thermodynamic integral equations 
for the antiferromagnetic ground state energy,
as originally observed by Lieb and Wu \cite{LW}. 

Our results strongly indicate that, sadly, wrapping interactions are 
{\it not} able to explain the three-loop discrepancies \cite{Call1,SS}
between gauge and string theory, as was originally hoped for 
in a proposal in \cite{BDS}. As discussed in chapter \ref{sec:bdsproof},
the Hubbard model simply does not seem to allow for two distinct ways 
to form the small BMN parameter $\lambda' \sim g^2/L^2$. Put differently,
in the Hubbard model there is no order of limits problem,
and wrappings just lead to ${\cal O}(g^{2 L})$ effects which 
disappear in the BMN limit.
This negative result seems to be in agreement with the complementary
findings in \cite{AJK}.

Actually, we cannot currently exclude that there might be
other, similar (modified, generalized Hubbard?) models which
also agree with BDS up to wrapping order, but differ from 
our current proposal in the wrapping terms. However, even if these
exist, we find it hard to believe that they will allow for a 
new way to form the BMN parameter $\lambda'$ at strong coupling.

These questions should be distinguished from the related, but distinct
(since the AdS/CFT discrepancy appears at three loops)
issue whether the BDS-Hubbard system is actually describing the gauge
theory's $\alg{su}(2)$ dilatation operator {\it at and beyond}
four-loop order. It is of course logically possible that the 
latter is {\it not} asymptotically given by the BDS chain at some loop
order larger than three. Assuming integrability, we would then
conclude that BMN scaling should break down at, or beyond,
four-loop order, cf.~\cite{BDS}. Then the BMN proposal 
\cite{BMN} along with the arguments of \cite{SZ} 
would be invalid for the gauge side. 

It should be clear from the preceding discussion that 
{\it we are in dire need of a perturbative
four-loop anomalous dimension computation in 
${\cal N}=4$ gauge field theory}. Of particular importance
would be the four-loop dimension of the Konishi field.
If it turns out to agree with our finding in this paper
($-318\,g^8$), our attempts to identify the $\alg{su}(2)$ sector
of the ${\cal N}=4$ dilatation operator with the Hubbard Hamiltonian
will, in our opinion, become very plausible. If it disagrees, the
search for the correct all-loop dilatation operator will have to be
continued.

Strong additional constraints come from considering the integrable
structure of the dilatation operator beyond the $\alg{su}(2)$ sector. 
The $\alg{su}(2)$ three-loop dilatation operator \cite{BKS} is 
naturally embedded in the maximally compact closed sector
$\alg{su}(2|3)$ \cite{dynamic}. 
The asymptotic BDS ansatz may also be lifted in a very
natural fashion to this larger sector \cite{S,BS2}. 
Here ``natural'' means that the ansatz (1) contains BDS as a limit,
(2) diagonalizes the three-loop dilatation operator in the 
$\alg{su}(2|3)$ sector, which is firmly established \cite{dynamic,EJS},
and (3) may be derived from a factorized S-matrix satisfying the
Yang-Baxter algebra \cite{S,BS2,Sproofs}. 
Actually, the asymptotic BDS ansatz may even be modified to include
non-compact sectors such as $\alg{sl}(2)$ \cite{S}, and lifted to
the {\it complete} theory \cite{BS2}, with symmetry $\alg{psu}(2,2|4)$.
Again, the construction seems compelling as it may be shown that
(1) the Bethe ansatz correctly diagonalizes to three loops twist-two 
operators \cite{S} whose dimensions are known form the work of 
\cite{MVV,KLOV},
(2) it also diagonalizes a twist-three operator to two loops 
which was confirmed using field theory in \cite{E}.
In fact, it may be proved (3) that it diagonalizes to two loops 
the dilatation operator in the $\alg{psu}(1,1|2)$ sector which 
has recently been computed by Zwiebel, using algebraic means in 
\cite{Z}, and (4) for $\alg{sl}(2)$ one may derive the ansatz at 
two loops directly from the field theory \cite{ES}.  
Finally, the entire $\alg{psu}(2,2|4)$ ansatz may again be derived from
an S-matrix satisfying the Yang-Baxter equation \cite{Sproofs}.
It is important to note that the structure of the S-matrix,
as well as, as a consequence, the nested asymptotic Bethe ansatz,
are nearly completely constrained by symmetry \cite{Sproofs}, up
to a global scattering ``dressing factor'' \cite{AFS,S,BS2}. 
This means that e.g.~the Inozemtsev model \cite{SS} is ruled out 
\cite{Sproofs} as an all-loop candidate. 
It also means that a possible breakdown of BMN scaling, confer the
discussion above, could only be caused by the dressing factor,
starting at or above four loops. See also \cite{BK}.
Incidentally, it would be very interesting to understand whether
short-range formulations also exist for other (or even all)
asymptotically integrable long-range spin chains \cite{stringchain},\cite{BK}.

From the preceding discussion we conclude that it will be crucial
to investigate whether the twisted Hubbard model may be extended to
sectors other than $\alg{su}(2)$, and eventually to the full symmetry
algebra $\alg{psu}(2,2|4)$. A further constraint will be that this
extension asymptotically yields the Bethe equations of \cite{BS2}. 
It would be exciting if finding the proper short-range formulation
of the full dilatation operator resulted, when restricting to
$\alg{su}(2)$, in a model that also asymptotically generates BDS but
differed from the specific Hubbard Hamiltonian we discussed in 
this paper. At any rate we find it likely, given the results of
this work, that such a {\it short-range} formulation of the gauge theory 
dilatation operator exists. It will be interesting to see whether
the latter also eliminates the length-changing operations
which appear in the current long-range formulation as a 
``dynamic'' spin chain \cite{dynamic}.

An intriguing if puzzling aspect of our formulation is that
the Hubbard model has many more states than the perturbative gauge
theory in the $\alg{su}(2)$ sector. For a length $L$ operator
we have roughly $2^L/L$ cyclic states in the spin chain and in the 
gauge theory, and $4^L/L$ cyclic states in the Hubbard model,
cf.~section \ref{sec:bdsproof1}. Is this an artifact of the
incompleteness, or erroneousness, of our identification, or
a first hint at a rich non-perturbative structure of planar
$\cN=4$ gauge theory? Does it possibly tell us that the 
fields appearing in the Lagrangian of the $\cN=4$ theory are
composites of more fundamental degrees of freedom (such as the
``electrons'' of our model)? 
A description of the dilatation operator in terms of fermionic and bosonic
degrees of freedom akin the fermionic variables in the Hubbard model,
which works to two loop order, was attempted in \cite{alday}.
Note also that the Hubbard model has a
second ``hidden'' $\alg{su}(2)$ symmetry \cite{books}. Our
twisting procedure actually breaks the symmetry through the
boundary conditions. Thermodynamically, however, i.e.~in the large $L$ limit,
the symmetry is still present.
The mechanism is reminiscent of the considerations of 
Minahan \cite{Joe}, but the details appear to be different.

Concerning the proposed AdS/CFT duality \cite{ads}, our result,
for the moment, just deepens the mystery of the vexing
``three-loop discrepancies'' of \cite{Call1},\cite{SS}.
The dual string theory is classically integrable \cite{BPR},
which leads to a complete solution of classical string
motions \cite{FT} in terms of an algebraic curve 
\cite{KMMZ}. The uncovered integrable structure is very similar 
\cite{AS} but different \cite{SS},\cite{BDS} from the one of the 
(thermodynamic limit of) gauge theory. 
Much evidence was found that the string theory is also 
{\it quantum} integrable. This can be established by a spectroscopic
analysis of the spectrum of strings in the near-BMN limit
\cite{Call1},\cite{Call2}, which shows that it may be
``phenomenologically'' explained by factorized scattering
\cite{AFS},\cite{S}. Again, the integrable structure is similar
but, at the moment, appears to differ. 
Some progress has also been made towards deriving quantum
integrability directly from the string sigma model \cite{AF}.

It would be exciting to find a Hubbard-type short-range model which 
reproduces the string theory results. Recently it was demonstrated
by Mann and Polchinski \cite{MP}
that conformal quantum sigma models can give Bethe equations whose classical
limit reproduces (in the $\alg{su}(2)$ sector) the bootstrap 
equations of \cite{KMMZ}. There is one structural feature of their approach
which strongly resembles the considerations in this paper:
In order to be able to treat the  $\alg{su}(2)$ case they need to
employ a {\it nested} Bethe ansatz, which is reminiscent of the
Lieb-Wu equations of the Hubbard model. A difference, however, is
that in their case elementary excitations of the same type are 
interacting with a non-trivial S-matrix, while in our model 
identical fermions are free.


\bigskip
\leftline{\bf Acknowledgments}

\noindent 
D.S.~and M.S.~thank the {\it Kavli Institute for Theoretical Physics},
Santa Barbara, for hospitality.
D.S. thanks the  {\it Albert-Einstein-Institut,} Potsdam, for hospitality.
We also thank Niklas Beisert, David Berenstein, Zvi Bern,  
Burkhard Eden, Sergey Frolov, Frank G\"ohmann, David Gross,
Volodya Kazakov, Ivan Kostov, 
Juan Maldacena, Nelia Mann, Jan Plefka, Joe Polchinski,  Radu Roiban and 
Aureliano Skirzewski
for their interest and for helpful discussions.
This research was supported in part by the National Science Foundation 
under Grant No. PHY99-07949. 


\appendix


\section{Effective Spin Hamiltonian: Perturbation Theory and 
Computation Schemes}
\label{app:effective}

\subsection{Perturbation Theory for Degenerate Systems}
\label{app:effective1}

Consider a system which is described by a Hamiltonian $H_0$. Assume
that the spectrum of $H_0$ is discrete, and that the system is in a stable 
state with energy $E^0_a$. In general the subspace $U_a$ 
corresponding to 
an eigenvalue $E^0_a$  has dimension  $g_a$, where $g_a$ 
is the degeneracy of the level
$E^0_a$.  Let us denote by 
$|u_1 \rangle,...,|u_{g_a}\rangle$ the vectors spanning $U_a$. 
What happens if we add a small interaction $+\lambda V$? 
In general we have a set of subspaces 
$\mathcal{E}_1,...,\mathcal{E}_n$, for which 
$\mathcal{E}_1 (\lambda)+...+\mathcal{E}_n (\lambda)\to U_a$ 
when $\lambda \to 0$ and dim$(\mathcal{E}_1+...+\mathcal{E}_n)=g_a$. 
If $\lambda$ is sufficiently small, we may assume that there exists 
a one-to-one correspondence between $U_a$ and 
$W=\mathcal{E}_1 (\lambda)+...+\mathcal{E}_n(\lambda)$. 
This correspondence is established by a transformation, to be found.
Let $|\phi\rangle$ be any state in the Hilbert space generated by $H_0$. 
Its projection on $U_a$ is formally realized by :
\begin{equation} \label{P0}
P_0=\frac{1}{2\pi i} \oint_{C_0} \frac{dz}{z-H_0}\, ,
\end{equation} 
where the contour $C_0$ is enclosing only the eigenvalue $E^0_a$ of $H_0$. 
From this discussion we conclude that the projector on the subspace $W$ is given by
\begin{equation} \label{PTeil1}
P=\frac{1}{2\pi i} \oint_{C} \frac{dz}{z-H_0-\lambda V}\, .
\end{equation}
The contour $C$ encloses the $n+1$ points $E^0_a$, 
$E_1(\lambda),...,E_n(\lambda)$ (the last $n$
collapse to $E^0_a$ when $\lambda \to 0$). Using the identity
\begin{equation}\label{trivialid}
\begin{array}{cccccc} \displaystyle \frac{1}{z-H_0
-\lambda V}=\frac{1}{z-H_0} (z-H_0+\lambda V-\lambda V) 
\frac{1}{z-H_0-\lambda V}=\frac{1}{z-H_0} \\  
\displaystyle 
+\frac{1}{z-H_0}\lambda V \frac{1}{z-H_0-\lambda V}\, ,\end{array}
\end{equation}
we immediately get the expansion
\begin{equation} \label{PTeil2}
P=\frac{1}{2 \pi i} \oint_{C} dz \frac{1}{z-H_0} \sum^{\infty}_{n=0} 
\lambda^n (V \frac{1}{z-H_0})^n\, .
\end{equation}
Careful use of the generalized Cauchy integral formula leads to the expansion
\begin{equation} \label{PTeil3}
P=P_0-\sum^{\infty}_{n=1} \lambda^n \sum_{k_1+...+k_{n+1}=n,\  k_i \geq 0} 
S^{k_1}V S^{k_2} V...VS^{k_{n+1}}\, ,
\end{equation}
where one defines
\begin{equation} \label{propagators}
S^0 \equiv -P_0, \qquad S^k=\biggl( (1-P_0)\frac{1}{E_0-H_0}\biggr) ^k 
\qquad \mbox{for} \quad k > 0\, .
\end{equation}
Naively one would expect that the correspondence between $U_a$ and $W$ is 
realized by the projector  $P$, i.e.~that any state $|\psi\rangle \in W$ 
can be written as
\begin{equation} \label{naivecorr}
|\psi\rangle=P |\phi\rangle\, ,
\end{equation}
where $|\phi\rangle$ is some vector in $U_a$. This would allow us to bring 
the eigenvalue problem
\begin{equation}\label{schrod}
H|\psi\rangle=E |\psi\rangle\, ,
\end{equation}
with
\begin{equation} \label{fullh}
H=H_0+\lambda V\, ,
\end{equation}
to the subspace $U_a$
\begin{equation}\label{eigenvp}
H P |\phi\rangle=E P|\phi\rangle\, .
\end{equation}
The disadvantage of this procedure is that $E P$ is not proportional to the identity
map. Furthermore $P$ does not preserve the norm of the states. 
The problem of this 
effective overlap has been solved by L\"owdin \cite{L}. One introduces 
renormalized states (which are still states from $U_a$)
\begin{equation} \label{renormkets}
|\hat{\phi}\rangle=(P_0 P P_0)^{1/2} |\phi\rangle\, ,
\end{equation}
and thus one is lead to the introduction of the $U_a \leftrightarrow W$ 
correspondence operator $\Gamma$:
\begin{equation} \label{gamma}
\Gamma=P P_0 (P_0 P P_0)^{-1/2}\, ,
\end{equation}
where
\begin{equation} \label{squareroot}
(P_0 P P_0)^{-1/2} \equiv P_0 + \sum^{\infty}_{n=1} \frac{1}{4^n} 
\left(\begin{array}{c} 2n \\ n \end{array} \right) 
\biggl[P_0 (P_0-P) P_0 \biggr]^n\, ,
\end{equation}
plus an analogous formula for $(P_0 P P_0)^{1/2}$. 
One can then prove that $\Gamma^{\dagger} \Gamma =P_0$, so
\begin{equation} \label{normcons}
(\Gamma |\phi\rangle, \Gamma |\phi'\rangle)=(|\phi\rangle,|\phi'\rangle)\, ,
\end{equation}
and the transformation preserves the norm. We may now substitute equation 
(\ref{schrod}) by an effective equation
\begin{equation} \label{effeigenvp}
(h-E)|\phi'\rangle=0\, ,
\end{equation}
where
\begin{equation} \label{heff}
 h \equiv \Gamma^{\dagger} H \Gamma\, .
\end{equation}

The operator $h$ is the {\it effective} Hamiltonian. To find it 
for the Hubbard model at half-filling we put
\begin{equation} \label{subs}
H_0=t U \sum_{i=1}^L c^{\dagger}_{i \uparrow} 
c_{i \uparrow} c^{\dagger}_{i \downarrow} c_{i \downarrow}, 
\qquad V=\sum_{i<j}^L t_{ij} (X_{ij}+X_{ji}), \qquad X_{ij}=
\sum_{\sigma=\uparrow, \downarrow} c^{\dagger}_{i \tau} c_{i \tau}\, ,
\end{equation}
 One may show 
that the odd powers disappear from the expansion of $h$, as one would 
expect from the 'hopping and hopping back' random walk interpretation:
\begin{equation} \label{evenpowerexp}
h=\lambda^2 h_2+\lambda^4 h_4+\lambda^6 h_6+...\, .
\end{equation}

\subsection{Computation Schemes}
\label{app:effective2}

Performing the computations can be divided into three stages:
\paragraph{Stage 1}
\vspace{6mm}
This stage consists of evaluating the effective Hamiltonian (\ref{heff}) 
to a given order. This is a tedious problem beyond the first few orders. 
One can however use 
that $E^0_a=0$ for the half-filled Hubbard model, whence in order to get 
$h$ to n-th order, one only needs to evaluate $\Gamma$ to $(n-1)$-th 
order. Furthermore it can be proved that any term of the form
\begin{equation} \label{vanishingterms}
P_0 V S^{k_1} V S^{k_2}...V S^{k_{n-1}} V P_0\, ,  \quad \qquad k_i \geq 1\, ,
\end{equation}
for odd $n$ vanishes identically. This two observations greatly speed up 
the calculations. A program in FORM (see \cite{V}) was written to 
perform this stage of the calculations.\\ 
The result we found up to three-loop reads 
\begin{equation}
\label{formal}
\begin{array}{c} h= + \lambda^2  ( P_0VSVP_0 )
+ \lambda^4  ( P_0VSVSVSVP_0  - \frac{1}{2} P_0VSVP_0VSSVP_0 \\ 
-\frac{1}{2}P_0VSSVP_0VSVP_0 )\\
+\lambda^6 \big(P_0VSVSVSVSVSVP_0 -  \frac{1}{2}P_0VSVSVSVP_0VSSVP_0  \\
-\frac{1}{2}P_0VSVSVSSVP_0VSVP_0
- \frac{1}{2}P_0VSVSSVSVP_0VSVP_0 \\ - \frac{1}{2}P_0VSVP_0VSVSVSSVP_0  
- \frac{1}{2}P_0VSVP_0VSVSSVSVP_0 \\
+\frac{1}{2}P_0VSVP_0VSVP_0VSSSVP_0-\frac{1}{2}P_0VSVP_0VSSVSVSVP_0  \\
+\frac{3}{8} P_0VSVP_0VSSVP_0VSSVP_0- \frac{1}{2}P_0VSSVSVSVP_0VSVP_0  \\
- \frac{1}{2}P_0VSSVP_0VSVSVSVP_0
+ \frac{1}{4}P_0VSSVP_0VSVP_0VSSVP_0 \\+ \frac{3}{8}P_0VSSVP_0VSSVP_0VSVP_0 
+ \frac{1}{2}P_0VSSSVP_0VSVP_0VSVP_0 \big)\, , \end{array}
\end{equation}
It is indeed
equivalent to the expansion obtained by Klein and Seitz in \cite{KS}.

\paragraph{Stage 2}
\vspace{6mm}
This stage consists of substituting (\ref{subs}) into $h_{2n}$ as calculated 
in stage 1. The process of substitution can be well visualized by assigning 
to each $X_{i j}$ an oriented line, starting at $j$ and ending in $i$, see
Fig.1a. Products of the $X$ operators are represented by an oriented set of 
arrows, with the understanding that the lowest lying arrow corresponds 
to the last operator in the product. A curly bracket around a set of arrows 
denotes a sum over different locations of the arrows.
One can interpret these diagrams as virtual displacements of spins.
It was proved that the perturbation expansion consists only of 
\textit{linked} diagrams (see \cite{KS} for details). 
Each diagram is multiplied by a suitable factor following from the structure 
of the $h_{2n}$ expansion.
\paragraph{Stage 3}
\vspace{6mm}
This final stage consists of evaluating the diagrams obtained in 
stage 2. Since the diagrams are closed, for each lattice site $i$ the number 
of arrows starting and ending at $i$ is the same. Keeping in mind the 
definition of $X_{ij}$, and using anti-commutation relations for every 
diagram connecting $r$ lattice sides, we can assign each diagram
linear combinations of terms of the form
\begin{equation} \label{terms}
N(i_{1},\tau_1,\tau_2)...N(i_{k},\tau_{2k-1},\tau_{2k}) \qquad  k\leq r\, ,
\end{equation}
where
\begin{equation} \label{Ndef}
N(i,\tau, \sigma)=c^{\dagger}_{i \tau} c_{i \sigma}\, .
\end{equation}
Furthermore one may rewrite each diagram in terms of spin components
by means of the relations
\begin{equation} \label{SplusSminus}
S^{+}_{i}=c^{\dagger}_{i \uparrow} c_{i \downarrow}\, , 
\qquad S^{-}_{i}=c^{\dagger}_{i \downarrow} c_{i \uparrow}\, ,
\end{equation}
and
\begin{equation} \label{Sz}
S^{z}_{i}=\frac{1}{2}(c^{\dagger}_{i \uparrow} c_{i \uparrow} 
-c^{\dagger}_{i \downarrow} c_{i \downarrow}) \ 
\simeq c^{\dagger}_{i \uparrow} c_{i \uparrow}-\frac{1}{2} \  
\simeq 1/2 -c^{\dagger}_{i \downarrow} c_{i \downarrow}\, ,
\end{equation}
where the last two equalities are only valid when acting on 
states with a single electron per site. This is however our case, after 
putting the diagrams into the form (\ref{terms}). The whole 
procedure is carried out in FORM.
\vspace{6 mm}
\begin{center}
\includegraphics[scale=0.5]{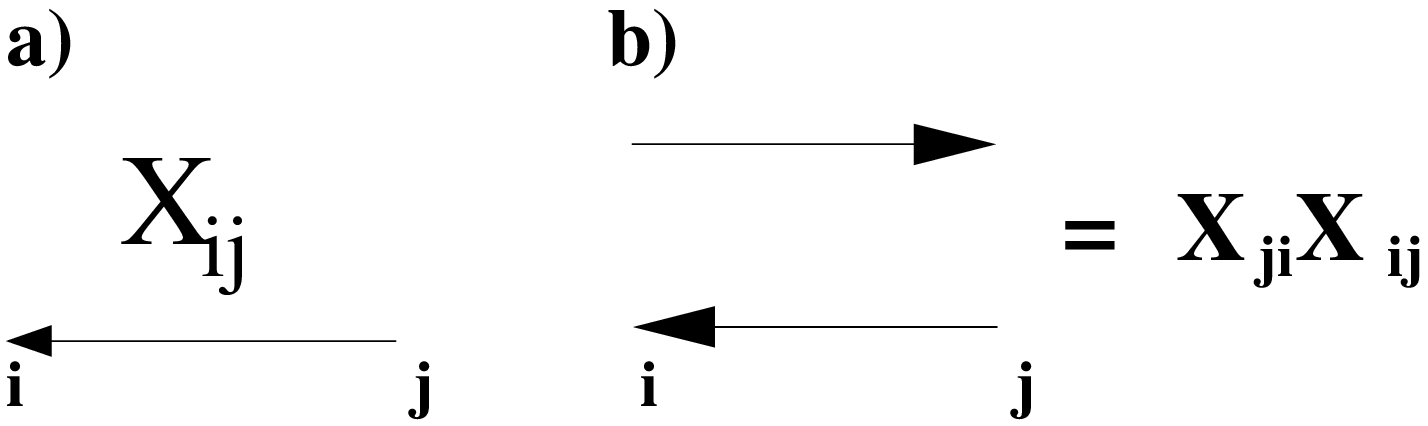}
\end{center}
\begin{center}
\small 
FIG. 1. a) To each $X_{i j}$ operator we assign an oriented arrow 
emerging from site j. b) A product of operators is represented by an
ordered set of arrows. The lowest lying arrow corresponds to the last 
operator in the product.
\end{center}

\subsection{Twist factors}
\label{app:effective3}

Equation (\ref{Hflux}) differs from (\ref{hubbard}) by the fact, that each 
hopping to the right is multiplied by a factor of $e^{i \phi}$, while 
hopping to the left gets an extra factor of $e^{-i \phi}$. Since the 
perturbation theory consists only of closed diagrams, we conclude that 
these factors cancel at the end.

This reasoning is generally true for long spin chains. A notable exception 
is when the chain is sufficiently short such that a spin can hop around the 
ring. This happens for example at two loops and $L=4$. There are two diagrams 
corresponding to this process. They are related to each other by
reversing all arrows in one of them. The two diagrams have thus 
weights differing by  factors of opposite signature
($e^{i \phi L}=i$ and $e^{-i\phi L}=-i$, c.f.~\eqref{twist}) and they 
therefore cancel each other. Thus putting the twist results in 
eliminating these unwanted demi-wrapping terms.


\section{\texttt{Mathematica} Code for the Perturbative 
Solution of the Lieb-Wu Equations}
\label{app:code}

In this appendix we will collect some \texttt{Mathematica} routines
which allow for the  immediate solution of the BDS equations 
\eqref{bds},\eqref{u},\eqref{bdseng} and the
Lieb-Wu equations \eqref{liebwu1},\eqref{liebwu2},\eqref{liebwueng}
for systems of relatively small lengths
$L$ once the one-loop solution is known. The necessary input is thus 
the collection of one-loop Bethe roots $\{u_k\}$, i.e.~the
solution of \eqref{XXX2} for the state in question.
The one-loop roots for the first few states may be found in appendix A of
\cite{BDS}. The below routines may therefore be used to check our
claims about the agreement (below wrapping order) and 
disagreement (at and beyond wrapping order) between the BDS ansatz 
and the Lieb-Wu ansatz on various specific states. There is 
however one restriction where the program does not
directly apply: There are a number of ``singular'' states
\cite{BMSZ,BDS} with three special unpaired one-loop roots 
$u_1=-\frac{i}{2}, u_2=0, u_3=\frac{i}{2}$ which require regularization.

These definitions set up the algorithm:

\begin{tabbing}
\verb"uu[k_, l_] := Sum[u[k, s]g^s, {s, 0, 2l - 2}];"
\\ \\
\verb"qq[n_, l_] := Sum[q[n, s]g^s, {s, 0, 2l - 2}];"
\\ \\
\verb"\[Phi][L_] := If[EvenQ[L] == True, Pi/( 2L), 0];"
\\ \\
\verb"x[u_] := u/2(1 + Sqrt[1 - 2 g^2/u^2])"
\\ \\
\verb"BDS[L_, M_, l_] :="
\\
\verb"Table[(x[uu[k, l] + I/2]/x[uu[k, l] - I/2])^L +"
\\
\verb"Product[(uu[k, l] - uu[j, l] + I)/(uu[k, l] - uu[j, l] - I),"
\\
\verb" {j, 1, M}], {k, 1, M}]"
\\ \\
\verb"EBDS[M_, l_] := Sum[I/x[uu[k, l] + I/2] - I/x[uu[k, l] - I/2], {k, M}]"
\\ \\
\verb"LW1[L_, M_, l_] :=Table[Exp[I qq[n, l] L] -Product["
\\
\verb"(uu[j, l] - Sqrt[2]g Sin[(qq[n, l] + \[Phi][L])] - I/2)/"
\\
\verb"(uu[j, l]-Sqrt[2]g Sin[(qq[n, l] + \[Phi][L])] + I/2),{j, M}], {n, L}]"
\\ \\
\verb"LW2[L_, M_, l_] := Table[Product["
\\ 
\verb"(uu[k, l] - Sqrt[2]g Sin[qq[n, l] + \[Phi][L]] + I/2)/"
\\ 
\verb" (uu[k, l] - Sqrt[2]g Sin[qq[n, l] + \[Phi][L]] - I/2),{n, L}] +"
\\ 
\verb"Product[(uu[k, l] - uu[j, l] + I)/(uu[k, l] - uu[j, l] - I),"
\\ 
\verb"{j, 1, M}], {k,1, M}]"
\\ \\
\verb"ELW[L_, l_] := Sqrt[2]/g Sum[Cos[(qq[n, l] + \[Phi][L])], {n, L}]"
\end{tabbing}

In order to find the prediction of the BDS chain for e.g.~the 
state with $L=5$ and $M=2$, where the two
one-loop Bethe roots are $u_1=\frac{1}{2}$ and $u_2=-\frac{1}{2}$,
we then compute, to e.g.~$l=8$ loops,

\begin{tabbing}
\verb"Clear[u]; Clear[q]; L = 5; M = 2; l = 8;"
\\ \\
\verb"u[1, 0] = 0.5; u[2, 0] = -0.5;"
\\ \\
\verb"Do[xxx = Chop[Series[BDS[L, M, 2l + 1], {g, 0, ll - 2}]];"
\\ 
\verb"yyy = Flatten[Chop[Solve[Coefficient[xxx, g, ll - 2] == 0]]];"
\\ 
\verb"Do[u[k, ll - 2] = yyy[[k]][[2]], {k, 1, M}], {ll, 3, 2 l + 1}];"
\\
\verb"Series[EBDS[M, 2l], {g, 0, 2l - 2}] // Chop // Rationalize"
\end{tabbing}

If we are, on the other hand interested in the correct result of
the Hubbard model, we compute instead

\begin{tabbing}
\verb"Clear[u]; Clear[q]; L = 5; M = 2; l = 8;"
\\ \\
\verb"u[1, 0] = 0.5; u[2, 0] = -0.5;"
\\ \\
\verb"Do[q[n, 0] = 2 Pi/L(n - 1), {n, 1, L}];"
\\ 
\verb"Do[xxx = Chop[Series[LW2[L, M, 2l + 1], {g, 0, ll - 2}]];"
\\ 
\verb"yyy = Flatten[Chop[Solve[Coefficient[xxx, g, ll - 2] == 0]]];"
\\
\verb"Do[u[k, ll - 2] = yyy[[k]][[2]], {k, 1, M}];"
\\
\verb"uuu = Chop[Series[LW1[L, M, 2l + 1], {g, 0, ll - 2}]];"
\\ 
\verb"vvv = Flatten[Chop[Solve[Coefficient[uuu, g, ll - 2] == 0]]];"
\\ 
\verb"Do[q[n, ll - 2] = vvv[[n]][[2]], {n, 1, L}], {ll, 3, 2 l + 1}];"
\\ 
\verb"Series[ELW[L, 2l + 1], {g, 0, 2l - 2}] // Chop // Rationalize"
\end{tabbing}


\section{Generic Twists} 
\label{app:generic}

In this appendix we study all the possible twisted boundary conditions 
for the Hubbard model
which are compatible with integrability and the way they affect the Lieb-Wu 
equations.
The results are essentially due to Yue and Deguchi \cite{YueDeguchi}, 
who studied the 
twisted boundary conditions associated to a model of two coupled $XY$ models 
which, upon a Jordan-Wigner transformation, is equivalent to the twisted 
Hubbard model.
Translating their results in terms of the Hubbard model, we obtain that the 
twists depend on 
six different constants
\begin{eqnarray}
\phi_\up=a_\up+N\;b_\up+M\;c_\up\\
\phi_\dn=a_\dn+N\;b_\dn+M\;c_\dn
\end{eqnarray}
while the corresponding version of the Lieb-Wu equations is
\begin{eqnarray}
\label{liebwu1twist}
&\ &e^{i\tilde q_nL}=\prod_{j=1}^M 
\frac{u_j-\sqrt{2}g\sin (\tilde q_n+\phi_\up) -i/2}
{u_j-\sqrt{2}g\sin  (\tilde q_n+\phi_\up)+i/2}\, ,
\qquad n=1,\ldots, N
\\ \nonumber
&\ &\prod_{n=1}^{N}
\frac{u_k-\sqrt{2}g\sin  (\tilde q_n+\phi_\up) +i/2}
{u_k-\sqrt{2}g\sin  (\tilde q_n+\phi_\up)- i/2}= e^{iL(\phi_\dn-\phi_\up)}
\prod_{\textstyle\atopfrac{j=1}{j\neq k}}^M
\frac{u_k-u_j +i}{u_k-u_j-i}\, .
\quad k=1,\ldots, M
\end{eqnarray}
The energy of the corresponding states is given by 
\begin{equation}
\label{liebwutwisteng}
E=\frac{\sqrt{2}}{g}\;
\sum_{n=1}^{N}\cos  ( \tilde q_n+\phi_\up)\;.
\end{equation}
After the duality transformation, the fermion number becomes $L-N+2M$, $g$ 
changes sign
and $\phi_\up\to \pi-\phi_\up$, and $\phi_\dn \to\phi_\dn$.
The dual Lieb-Wu equations are, for  generic twist
\begin{eqnarray}
\label{liebwu1twistdual}
&\ &e^{i q_nL}=\prod_{j=1}^M 
\frac{u_j-\sqrt{2}g\sin ( q_n-\phi_\up) -i/2}
{u_j-\sqrt{2}g\sin  (q_n-\phi_\up)+i/2}\, ,
\qquad n=1,\ldots, L-N+2M
\\ \nonumber
&\ &\prod_{n=1}^{L-N+2M}
\frac{u_k-\sqrt{2}g\sin  ( q_n-\phi_\up) +i/2}
{u_k-\sqrt{2}g\sin  ( q_n-\phi_\up)- i/2}= e^{iL(\phi_\dn+\phi_\up-\pi)}
\prod_{\textstyle\atopfrac{j=1}{j\neq k}}^M
\frac{u_k-u_j +i}{u_k-u_j-i}\, .
\quad k=1,\ldots, M
\end{eqnarray}
while the energy is
\begin{equation}
\label{liebwutwistengdual}
E=-\frac{M}{g^2}-\frac{\sqrt{2}}{g}\;
\sum_{n=1}^{L-N+2M}\cos  ( q_n-\phi_\up)\;.
\end{equation}
To obtain the BDS ansatz, the following conditions on the twists have to 
be satisfied
\begin{equation}
e^{iL(2\phi_\up-\pi)}=e^{iL(\phi_\dn+\phi_\up-\pi)}=-1\;,\quad {\rm or} 
\quad \phi_\up=\phi_\dn=\frac{\pi(L+1)}{2L}
\  {\rm mod}\  \frac{\pi}{L}\;.
\end{equation}
These are exactly the values we used in \eqref{twist}, so we infer that 
there is no other possibility
to choose the twists compatible with the BDS ansatz.

\section{Further Details on the One-Magnon Problem}
\label{app:one-magnon}

In section \ref{sec:bdsproof1} we discussed how to account for
all states of the twisted Hamiltonian acting on $L-1$ up spins and
$M=1$ down spin. Recall that in the Hubbard model this corresponds
to a {\it two-body} problem, hence there are $L^2$ states.
$L$ of these states are bound states, whose dispersion law 
\eqref{dispersion} coincides
with the one of the magnons in the BDS chain. This law turns,
using $p=2 \pi\, n/L$,
into the BMN square-root formula
\begin{equation}
\label{bmn}
g^2\,E \simeq -1+\sqrt{1+\lambda'\, n^2}\,
\end{equation}
if we scale $\lambda =8\, \pi^2\, g^2\rightarrow \infty$, 
$L \rightarrow \infty$
while holding  $\lambda'=\lambda/L^2$ fixed.
As we showed in sections \ref{sec:bdsproof2} and \ref{sec:bdsproof3}, 
the scattering of these bound states is as in the near-BMN limit
of the BDS chain. It is therefore, at third order $(\lambda')^3$,
incompatible with the predictions of string theory \cite{Call1}.

One potential way out of this trouble would be to find 
other states in our model which scatter as in string theory.
A prerequisite is that the coupling constant dependence of 
the dispersion law of such candidate states is again as in
\eqref{bmn}, with, possibly, a different constant part.  
In particular, among the real solutions we identified in
section \ref{sec:bdsproof1}, there were states of exact dimension 
$\Delta=L-1$
which resembled ``BPS states''. Let us therefore work out
the dispersion law of the nearby ``near-BPS'' states.
This requires studying the solutions of \eqref{sinbeta}
for small $q=\pi/L\, n$ and large $g$. Expressing \eqref{sinbeta}
through the BMN coupling $\lambda'$, we find
\begin{equation}
\label{sinbeta2}
\sin\left(b\right)=
\frac{1}{\sqrt{\lambda'}\,n}\,
\tan\left(b\,L\right)\, .
\end{equation}
Since this equation should hold as $\lambda' \rightarrow 0$,
we recover the $L-1$ mode numbers $m$ \eqref{bmodes}:
\begin{equation}
\label{bexpansion}
b=\frac{\pi}{L}\,m+\frac{\delta b_m}{L}\, .
\end{equation}
Now, $\delta b_m$ should be at most of order $\cO(1)$, i.e.~it 
should not be too large so as to move out of the branch of
$\tan(b_m\,L)$ defined by \eqref{bexpansion}, and should tend
to zero if $\sqrt{\lambda'}\, n\rightarrow 0$. This yields from
\eqref{sinbeta2} 
$\delta b_m \simeq \sqrt{\lambda'}\, n\,\sin(\frac{\pi}{L}\,m)$.
Substitution into the expression for the energy of the real solutions
\begin{equation}
\label{realdispersion}
E=-\frac{1}{g^2}+\frac{2 \sqrt{2}}{g}\,\sin\left(q\right)\,
\cos (b)
\end{equation}
gives for the dimension $\Delta$
\begin{equation}
\label{realdispersion2}
\Delta = L-1+\sqrt{\lambda'}\,n \cos(\frac{\pi}{L}\,m)
-\frac{1}{L}\,\lambda'\,n^2\,\sin^2\left(\frac{\pi}{L}\,m\right)
+{\cal O}(1/L)\;.
\end{equation}
We see that we generically lift the $L-1$ degenerate ``BPS-states''
with a term non-analytic in $\lambda'$. If we concentrate on
mode numbers close to $m\simeq L/2$ we can suppress the non-analytic
$\sqrt{\lambda'}$ term. The next term is then analytic in
$\lambda'$, but {\it subleading} in $1/L$. 
It is interesting to note that there is a possibility to reproduce
a BMN-like dispersion relation, by choosing 
$m$ such that
\begin{equation}
\cos(\frac{\pi}{L}\,m)=\frac{1}{2} \sqrt{\lambda'} n+{\cal O}(1/L)\;,
\end{equation}
so that the conformal dimension 
would be analytic in $\lambda'$ up to terms of order $1/L$
\begin{equation}
\label{realdispersion3}
\Delta = L-1+\frac{1}{2} \lambda' n ^2+{\cal O}(1/L)\;.
\end{equation}
However, such a choice for $m$ is not continuous in $\lambda'$ and
cannot be sensibly interpreted in terms of  BMN states.
The ``BPS-states'' we found are thus very different from the usual
ones, 
and the BMN states may not be expected to hide
among the continuum of real solutions.

Finally note that any one of the $L$ bound states of section 
\ref{sec:bdsproof1} can disappear\footnote{
This is the so-called ``redistribution phenomenon'' \cite{EsslerFinite} 
and is responsible for rendering all the fermion momenta real in the 
extreme $g$ limit, $g\gg L$, which corresponds to the  free fermion limit.
}
if $g$ is very close to $L$.
One may show that in this case a further real solution with mode number
$m=0$, which generically does not correspond to a solution of 
\eqref{sinbeta}, appears.
Unfortunately this deconfinement phenomenon is 
also not suitable for finding the BMN states of string theory \cite{BMN},
as we are then not allowed to make the parameter $\lambda'$ in
\eqref{sinbeta2} arbitrarily small.

\section{Alternative Proof of the BDS Equations}
\label{app:alternative}

In this appendix we give an alternative proof to the BDS ansatz, using the 
original Lieb-Wu equation,
with a macroscopic number of fermionic excitations. As in the original paper 
\cite {LW}, we suppose that the fermion momenta are all real and they form, 
in the continuous limit,  a continuous density.
This proof is less effective than the one which starts from the dual Bethe 
ansatz, in the sense that
the finite size corrections are not under control, and the effect of the 
boundary conditions (twist)
is lost. However, it is interesting to see that the BDS equations are 
already contained in the 
integral equations of Lieb and Wu  \cite {LW}.

At half-filling, the Lieb-Wu equations can be written in the logarithmic 
form as
\begin{eqnarray}
\label{LWdens}
&\ &q_n=\phi+\frac{2\pi n}{L}-\frac{2}{L}\sum_{j=1}^M 
\arctan\frac{1}{2(u_j-\sqrt{2}
g\sin{q_n})}\;, \ \ n=1,\ldots,L\\
\label{LWdens2}
&\ &2\sum_{n=1}^L \arctan\frac{1}{2(u_k-\sqrt{2}
g\sin{q_n})}=2\pi m +2\sum_{\textstyle\atopfrac{j=1}{j\neq k}}^M 
\arctan\frac{1}{2(u_k-u_j)}\;.
\end{eqnarray}
The choice of the branch of the logarithm in \eqref{LWdens} is made 
by continuity, 
such that at $g=0$  there is exactly one electron per level.
For simplicity, we have remove the tilde on the variables $q_n$ and shifted 
them by
$\phi$. 
Taking the derivative of the first equation with respect to $q$  and defining 
the density $\rho(q)=(dn/dq)/L$ we obtain an equation for the density
\begin{equation}
\label{contdens}
2\pi \rho(q)=1+\frac{2}{L}\sum_{j=1}^M \;\frac{2\sqrt{2}g\cos{q}}{4(u_j
-\sqrt{2}g\sin{q})^2+1}\;.
\end{equation}
Our purpose is to study the case of a finite (arbitrary) number of magnons, so
we do not introduce a density for the magnons. 
instead, we evaluate the
left hand side of the second equation Lieb-Wu equation \eqref{LWdens2}
\begin{equation}
I(u_j)=2L\int_{-\pi}^\pi dq \;\rho(q) \arctan \frac{1}{2(u_j-\sqrt{2}
g\sin{q})}
\end{equation}
The second term in the density does not contribute to the integral $I(u)$. 
To compute the integral $I(u)$, we first take its derivative with respect
to $u$, so that the  cuts of the integrand disappear
\begin{equation}
\frac{d}{du}I(u)=\frac{L}{2\pi i}\int_{-\pi}^\pi dq \; \left(\frac{1}{u+i/2
-\sqrt{2}
g\sin{q}}-\frac{1}{u-i/2-\sqrt{2}
g\sin{q}}\right)\;.
\end{equation}
The integral over $q$ can be traded to a contour integral by a change of 
variable
$z=\sqrt{2}g\sin q$
\begin{equation}
\frac{d}{du}I(u)=\frac{L}{2\pi i}\oint _C \frac{i dz}{\sqrt{z^2-2g^2}} \; 
\left(\frac{1}{u^+-z}-\frac{1}{u^--z}\right)\;,
\end{equation}
where $C$ is the contour encircling the interval $[-\sqrt{2}g,\sqrt{2}g]$
clockwise. The contour $C$ cannot be shrunk to zero because of the 
obstruction created by the
square root in the integrand. The integral vanishes on the contour at 
infinity, so 
we can deform the contour
$C$ into two contours $C_+$ and $C_-$ which encircle the points $u^+$ and 
$u^-$ counterclockwise.
We obtain 
\begin{equation}
\frac{d}{du}I(u)=-i{L}\left(\frac{1}{\sqrt{u^+-2g^2}}
-\frac{1}{\sqrt{u^--2g^2}}\right)
=-i{L}\;\frac{d}{du\;}\ln \;\frac{x(u^+)}{x(u^-)}\;.
\end{equation}
The constant of integration can be easily seen to be zero, since $I(\infty)=0$.
Finally, the second Lieb-Wu equation 
\eqref{LWdens2} takes the form
\begin{equation}
\label{BDSbis}
\left(\frac{x^+(u_k)}{x^-(u_k)}\right)^L=
\prod_{\textstyle\atopfrac{j=1}{j\neq k}}^M
\frac{u_k-u_j +i}{u_k-u_j-i}\, .
\end{equation}
The magnon energy can be computed by the same means. In this case, only 
the second
term in  the density \eqref{contdens} contributes
\begin{equation}
E=\frac{\sqrt{2}}{g}L\int_{-\pi}^\pi dq\; \rho(q) \cos(q)=
\frac{\sqrt{2}}{\pi g}\sum_{j=1}^M
\int_{-\pi}^\pi dq\; \frac{2\sqrt{2}g\cos^2{q}}{4(u_j-\sqrt{2}g\sin{q})^2+1}\;.
\end{equation}
Again, the integral can be converted into a contour integral around the 
same contour $C$
which encircles the cut $[-\sqrt{2}g,\sqrt{2}g]$ clockwise
\begin{equation}
E=\frac{1}{ g^2} \sum_{j=1}^M\; \oint_C \frac{dz}{2\pi i}\;
\frac{ \sqrt{z^2-2g^2}}{(z-u_j^+)(z-u_j^-)}\;.
\end{equation}
As such, the integral does not vanish on the contour at infinity,
but we can freely add  to it a term which is regular across the cut and which 
removes the contribution from infinity
\begin{eqnarray}
\label{enBDSbis}
&\ &E=\frac{1}{ g^2}\sum_{j=1}^M
\oint_C \frac{dz}{2\pi i}\;\frac{ (\sqrt{z^2-2g^2}-z)}{(z-u_j^+)(z-u_j^-)}\\
&=&-\sum_{j=1}^M\oint_C \frac{dz}{2\pi i}\frac{x^{-1}(z)}{(z-u_j^+)(z-u_j^-)}=
i\sum_{j=1}^M\left(\frac{1}{x(u^+_j)}- 
\frac{1}{x(u^-_j)} \right)\;.\nonumber
\end{eqnarray}
Of course, the reader recognizes  \eqref{BDSbis} and \eqref{enBDSbis}
as the equations of BDS ansatz.



\end{document}